\documentclass{JHEP3} 

\usepackage{graphicx}
\usepackage{amsmath, amsfonts}
\usepackage{epsfig}

\newcommand{\non}{\nonumber\\}

\newcommand{\be}{\begin{equation}}
\newcommand{\ee}{\end{equation}}
\newcommand{\bea}{\begin{eqnarray}}
\newcommand{\eea}{\end{eqnarray}}
\newcommand{\ba}[1]{\begin{array}{#1}}
\newcommand{\ea}{\end{array}}

\title{Inverse magnetic catalysis in dense holographic matter}

\author{Florian Preis, Anton Rebhan, Andreas Schmitt\\
Institut f\"{u}r Theoretische Physik, Technische Universit\"{a}t Wien, 1040 Vienna, Austria\\ 
\\
\email{fpreis@hep.itp.tuwien.ac.at}\\
 \email{rebhana@hep.itp.tuwien.ac.at}\\
 \email{aschmitt@hep.itp.tuwien.ac.at}
  }

\abstract{We study the chiral phase transition in a magnetic field at finite temperature and chemical potential within
the Sakai-Sugimoto model, a holographic top-down approach to (large-$N_c$) QCD.
We consider the limit of a 
small separation of the flavor D8-branes, which corresponds to a dual field theory comparable to a Nambu-Jona Lasinio (NJL) model.
Mapping out the surface of the chiral phase transition
in the parameter space of magnetic field strength, 
quark chemical potential, and temperature, we find that 
for small temperatures the
addition of a magnetic field decreases the critical
chemical potential for chiral symmetry restoration -- in contrast
to the case of vanishing chemical potential
where, in accordance with the familiar phenomenon of magnetic catalysis,
the magnetic field favors the chirally broken phase.
This ``inverse magnetic catalysis'' (IMC)
appears to be associated with a previously found magnetic phase transition
within the chirally symmetric phase that
shows an intriguing similarity to a transition into the lowest Landau level. 
We estimate IMC to persist up to $10^{19}\, {\rm G}$ at low temperatures.
}

\keywords{Gauge-gravity correspondence, QCD, Chiral Lagrangians}

\begin{document}

\section{Introduction}
\label{sec:intro}

Spontaneous breaking of chiral symmetry in a system of relativistic fermions is profoundly affected by an external magnetic field.
A sufficiently strong, homogeneous magnetic field results in an effective dimensional reduction of the dynamics of the system. 
As a consequence, an instability with respect to condensation of fermion-antifermion pairs, i.e., with respect to the formation 
of a chiral condensate, occurs even at arbitrarily weak coupling. This is analogous to the Bardeen-Cooper-Schrieffer (BCS) mechanism of 
fermion-fermion pairing in a superconductor, where the effective dimensional reduction is achieved by the presence of a Fermi surface. 
The enhancing effect of the magnetic field on chiral symmetry breaking has been termed magnetic catalysis (MC) and has originally been discussed in 
Gross-Neveu \cite{Klimenko:1990rh,Klimenko:1992ch} and Nambu-Jona-Lasinio (NJL)  
\cite{Gusynin:1994re,Gusynin:1994va,Gusynin:1994xp} models and in QED \cite{Gusynin:1995gt}. Recently, 
this effect has also been reproduced in holographic models
with flavor branes subjected to magnetic fields
\cite{Filev:2007gb,Erdmenger:2007bn,Filev:2009xp,Filev:2010pm}.

In this paper, we consider  
the chiral phase transition under the influence of a magnetic field $B$ at 
finite temperature $T$ and 
quark chemical potential $\mu$. 
Our main interest is
QCD where chiral symmetry is spontaneously broken for sufficiently small 
temperatures and chemical potentials,
and where the effects of strong magnetic fields may be observable
in the chiral transition at small $\mu$ and large $T$
(namely in ultrarelativistic heavy-ion collisions) and
also at small $T$ and large $\mu$ in astrophysical systems (compact stars).
Indeed, it has been argued that extremely strong magnetic
fields of up to $\sim 10^{18}\, {\rm G}$ occur in non-central heavy-ion collisions \cite{Skokov:2009qp} and up to $\sim 10^{15}\, {\rm G}$ at the surface of magnetars  \cite{Duncan:1992hi} (possibly even up to $\sim  10^{19}\, {\rm G}$ 
in the interior \cite{Lai}). 
Our results may also be relevant for graphene \cite{graphene} 
which is under much better experimental control. Fermion excitations in 
graphene are effectively relativistic and MC manifests itself in a nonzero Dirac mass induced by electron-hole pairing 
\cite{Gusynin:2006gn,Herbut:2006cs,Gorbar:2008hu}, analogous to the constituent quark mass induced by quark-antiquark and quark-hole 
pairing in the QCD context.

In both QCD and condensed matter contexts it is important to develop a strong-coupling description of MC. 
To this end we employ the 
AdS/CFT correspondence \cite{Maldacena:1997re,Gubser:1998bc,Witten:1998qj,Witten:1998zw}, more precisely the Sakai-Sugimoto model 
\cite{Sakai:2004cn,Sakai:2005yt}. This holographic model, based on type-IIA string theory, is, in a certain (albeit 
inaccessible) 
limit dual to large-$N_c$ QCD. In contrast to most other holographic models, it accounts for the full chiral symmetry group by 
realizing left- and right-handed massless fermions through $N_f$ D8- and $\overline{\rm D8}$-branes in a background of $N_c$ D4-branes. Moreover, the model
has a confined and a deconfined phase, realized by two different background geometries. In the original version of the model, where the 
D8- and $\overline{\rm D8}$-branes  
are maximally separated in a compact extra dimension, the deconfinement and chiral phase transitions are identical and happen at a certain 
$T$ for all values of $\mu$ and $B$ (provided that any backreaction on the background is neglected). Here we are interested in a different 
limit of the model where the distance of the flavor branes is small and where a much richer phase structure is obtained. 
This limit can be understood as the NJL limit of the model \cite{Antonyan:2006vw,Davis:2007ka,Edalati:2009xc}.

The NJL model in its original form approximates the fermionic interaction by a point-like four-fermion interaction. 
It has been employed for the chiral 
phase transition in the presence of a background magnetic field at finite $\mu$ and/or $T$ in 
refs.\ \cite{Ebert:1999ht,Ebert:2003yk,Inagaki:2003yi,Boomsma:2009yk,Frolov:2010wn,Fayazbakhsh:2010bh,Chatterjee:2011ry}. We shall find
a phase diagram which shows a striking qualitative resemblance with some of the NJL results. In particular, we shall discuss that
at finite chemical potential and not too large 
magnetic field the chirally broken phase becomes disfavored by increasing the magnetic field, 
in stark contrast to MC. We term this effect inverse magnetic catalysis (IMC) and present a simple physical explanation, employing the
analogy to superconductivity.
Moreover, we shall discuss a discontinuity  in the quark density \cite{Lifschytz:2009sz} for small temperatures in the chirally symmetric phase. Our result for this 
discontinuity, in particular the comparison of its location with respect to the chiral phase transition to recent NJL results, 
supports its interpretation 
as a transition to the lowest Landau level. This is remarkable since 
with the exception of the Sakai-Sugimoto model \cite{Lifschytz:2009sz,Thompson:2008qw}, Landau-level-like structures have been discussed
in the AdS/CFT literature 
only in bottom-up scenarios \cite{Basu:2009qz,Denef:2009yy,Albash:2009wz,Albash:2010yr}. 
Here we shall see that the top-down approach of the Sakai-Sugimoto model 
suggests the presence of a lowest Landau level, but no further de Haas-van Alphen oscillations from higher Landau levels.

Our calculation builds upon previous work within the Sakai-Sugimoto model in the presence of a magnetic field 
\cite{Thompson:2008qw,Bergman:2008qv,Rebhan:2008ur,Rebhan:2009vc}, and generalizes the results for the chiral phase transition
in the $T$-$\mu$ plane at $B=0$ \cite{Horigome:2006xu} and the $T$-$B$ plane at $\mu=0$ \cite{Bergman:2008sg,Johnson:2008vna} to the 
entire $T$-$\mu$-$B$ space. (For recent discussions of the chiral phase transition in a magnetic field within other holographic models
see for instance refs.\ \cite{Evans:2010iy,Jensen:2010vd,Jensen:2010ga,Evans:2010hi}.)

The remainder of the paper is organized as follows. In sec.\ \ref{sec:setup} we explain the geometry of our holographic setup, introduce our 
notation, and derive the on-shell action and the equations of motion. In secs.\ \ref{sec:broken} and \ref{sec:symmetric} we treat the chirally 
broken 
and symmetric phases separately and point out discontinuities in the quark density in both phases. The main part of the paper is 
sec.\ \ref{sec:transition} where the chiral phase transition is discussed. After discussing several limit cases in sec.\ \ref{sec:B0}
-- \ref{sec:mu0} (including a discussion of the split of chiral and deconfinement transitions in sec.\ \ref{sec:mu0}) we
present our main results in sec.\ \ref{sec:IMC} before giving our conclusions in sec.\ \ref{sec:conclusions}.

\section{General setup}
\label{sec:setup}

\subsection{Context and brief summary of the model}

We consider the Sakai-Sugimoto model for one flavor, $N_f=1$, in the deconfined phase. 
The corresponding background geometry is given by the ten-dimensional supergravity description of $N_c$ D4-branes in type-IIA string theory
compactified on a supersymmetry breaking Kaluza-Klein 
circle \cite{Witten:1998zw}. 
Fundamental flavor degrees of freedom are implemented by $N_f$ D8- and 
$\overline{\rm D8}$-branes which are separated 
asymptotically by a given distance $L$ in the compactified dimension \cite{Sakai:2004cn,Sakai:2005yt}. 
Employing the probe brane approximation 
$N_c\gg N_f$, the background geometry will  
be fixed throughout the paper, while two qualitatively different embeddings of the flavor 
branes account for 
the chirally broken and chirally symmetric phases. The $U(N_f)$ gauge symmetries on the D8- and 
$\overline{\rm D8}$-branes are interpreted as left- and right-handed global symmetries of the dual field theory which lives at the 
4+1-dimensional boundary of the ten-dimensional space (including the compact extra dimension, which needs to be small to arrive at an 
effectively 3+1-dimensional field theory). In the case of disconnected flavor branes the system is invariant under
the full chiral group $U(N_f)_L\times U(N_f)_R$ while connected branes in the bulk 
lead to the smaller symmetry group $U(N_f)_{L+R}$, see fig.\ \ref{figcylinders}. 
This reflects the usual spontaneous chiral symmetry breaking of massless quarks. 

\FIGURE[t]{\label{figcylinders}
{ \centerline{\def\epsfsize#1#2{0.5#1}
\epsfbox{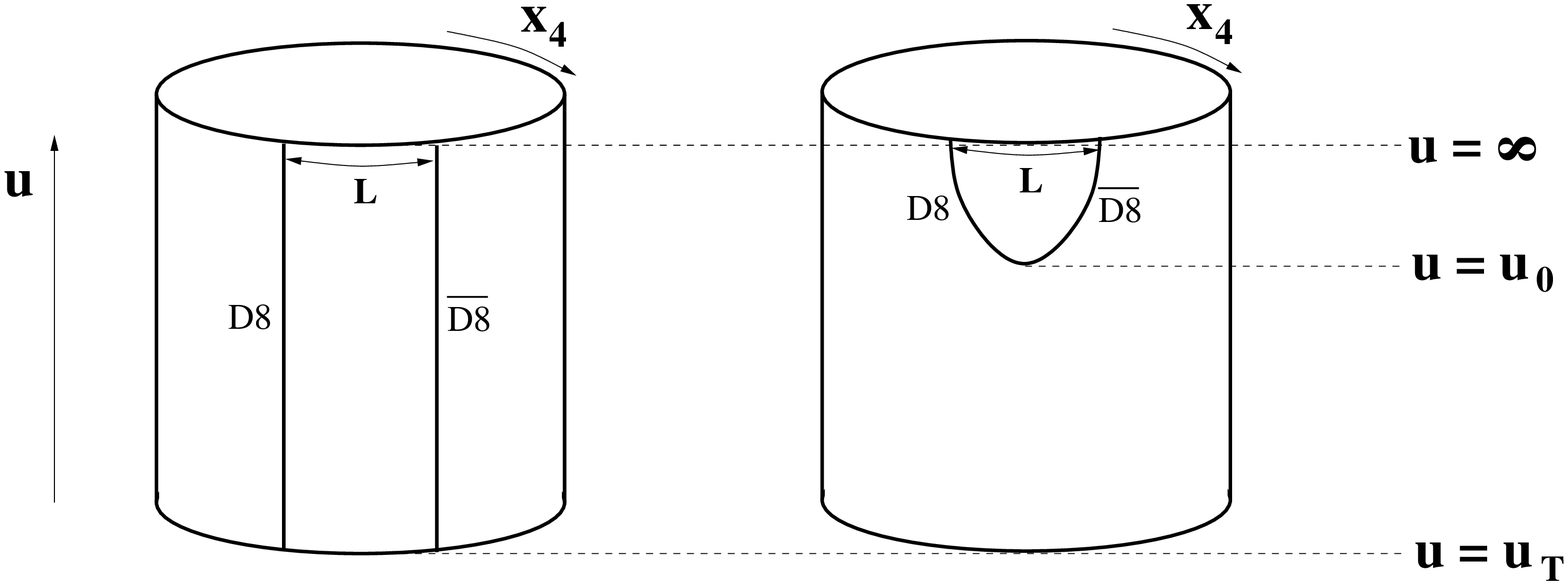}}
}
\caption{Schematic picture of the chirally symmetric (left) and chirally broken (right) phases in the deconfined phase of the Sakai-Sugimoto model 
(in the confined phase, the $x_4$-$u$ subspace is cigar-shaped, not cylinder-shaped). The $x_4$-direction is compactified on a circle
with radius $M_{\rm KK}^{-1}$, and $u$ is the holographic coordinate with $u=\infty$ being the boundary where the dual field theory lives. 
If the distance $L$ between the flavor branes is sufficiently small, a deconfined, chirally broken phase (right figure) becomes possible.
In this phase, the connected flavor branes are embedded nontrivially in the background according to a function $x_4(u)$ which has to 
be determined from the equations of motion which couple $x_4$ to the gauge fields on the flavor branes. The location of the tip of the 
connected branes $u_0$ is part of this solution. In this paper, we work in the ``NJL limit'' of the model, where $u_0\gg u_T$, i.e., 
the tip of the branes is far away from the horizon. For a fixed temperature $T$ -- which fixes $u_T$ --
the distance between $u_0$ and $u_T$ can be made arbitrarily large by choosing a sufficiently small separation $L$.  
 }  
}

In the original applications of the model the separation of the flavor branes $L$ is maximal, $L=\pi/M_{\rm KK}$, i.e., the D8- and 
$\overline{\rm D8}$-branes are put on opposite ends of the circle with radius $M_{\rm KK}^{-1}$. Here, $M_{\rm KK}$ is the Kaluza-Klein
mass which sets the mass scale below which adjoint scalars and fermions decouple from the dynamics of the dual field theory. 
For this maximal separation of flavor branes chiral symmetry 
is broken if and only if the system is confined. In other words, the chiral phase transition is dictated entirely by the 
background geometry. As a consequence, in the probe brane approximation the chiral transition is unaffected by all quantities that live on the 
flavor branes such as chemical potential and magnetic field.  
As an alternative to going beyond the probe brane approximation -- which is very difficult --,
such a ``rigid'' behavior can be softened by 
choosing a smaller separation of the flavor branes. This leads to a much richer phase structure, in particular a decoupling of the chiral and 
deconfinement phase transitions becomes possible. In fact, this decoupling
is realized for values of $L$ below a critical value $L\lesssim 0.3\pi/M_{\rm KK}$ \cite{Aharony:2006da}, yielding a deconfined, chirally broken 
phase for sufficiently small chemical potentials \cite{Horigome:2006xu}. Now that both connected and disconnected flavor branes are 
possible solutions in the deconfined background geometry, also a magnetic field affects the chiral transition. It has been shown 
that for vanishing chemical potential, the critical temperature above which chiral symmetry is restored increases with increasing 
magnetic field \cite{Bergman:2008sg,Johnson:2008vna}, in accordance with expectations from MC, as explained in the introduction. A simple 
consequence is that, in a certain regime of non-maximal separations $L$, a magnetic field may induce a splitting of chiral and deconfinement 
phase transitions. We discuss this effect in more detail in sec.\ \ref{sec:mu0}.
Such a splitting has been observed in a linear sigma model coupled to quarks and Polyakov loop 
\cite{Mizher:2010zb} and an NJL model with Polyakov loop (PNJL) \cite{Gatto:2010qs} (see however ref.\ \cite{Gatto:2010pt}), but 
has not been seen in lattice QCD calculations \cite{D'Elia:2010nq}.

For a large separation of the flavor branes, i.e., $L$ of the order of $\pi/M_{\rm KK}$, 
the main features of the phase diagram in the $T$-$\mu$ plane are the same as  
for large-$N_c$ QCD \cite{McLerran:2007qj}.
In the opposite limit $L\ll \pi/M_{\rm KK}$ the connected flavor branes 
are far away from the horizon so that the effect of confinement becomes less ``visible'' for the fundamental fermions.
Hence we expect this limit 
to correspond to a field theory where the dynamics of chiral symmetry breaking decouples from that of the gluons. Therefore, 
by varying the distance $L$ the Sakai-Sugimoto model 
interpolates between large-$N_c$ QCD ($L \sim \pi/M_{\rm KK}$) and a (non-local) NJL model ($L\ll \pi/M_{\rm KK}$) 
where there are no gluons and no confinement \cite{Antonyan:2006vw,Davis:2007ka}. 
Since in the former limit the $N_c^2$ many gluons dominate the phase diagram, 
the latter may in fact be an interesting limit for QCD at {\em finite} $N_c$,
at least at comparatively low temperatures and high quark number densitites.
Our main results correspond to the latter limit,
which indeed will show many similarities to those obtained
recently in NJL model calculations.

\subsection{Action on the flavor branes and equations of motion}

In the deconfined phase the induced (Euclidean) metric on the D8-branes is 
\bea
ds^2 &=&  \left(\frac{U}{R}\right)^{3/2}\Big[f(U)d\tau^2+\delta_{ij}dx^idx^j\Big] \non[2ex] 
&&+\, \left(\frac{R}{U}\right)^{3/2}\left\{\left[\frac{1}{f(U)}+\left(\frac{U}{R}\right)^3(\partial_U X_4)^2\right]dU^2+U^2d\Omega_4^2\right\} \, , 
\eea
where $(\tau,x_1,x_2,x_3)$ are the coordinates of 3+1-dimensional space-time, $X_4\in [0,2\pi/M_{\rm KK}]$ is the coordinate of the 
compactified extra dimension, $R$ is the curvature radius of the background, and $d\Omega_4^2$ is the metric of a four-sphere. The 
holographic coordinate on the flavor branes is denoted by $U$ with $U\in[U_T,\infty]$ (symmetric phase) and $U\in[U_0,\infty]$
(broken phase), see fig.\ \ref{figcylinders}, and  
\be
f(U) \equiv 1-\frac{U_T^3}{U^3} \, , \qquad U_T \equiv \left(\frac{4\pi}{3}\right)^2T^2R^3
\ee
with the temperature $T$. We have used capital letters for the coordinates $X_4$, $U$ to reserve lower-case letters for their dimensionless
versions introduced below.  

The action on the D8-branes has a Dirac-Born-Infeld (DBI) and Chern-Simons (CS) part,
\be
S = S_{\rm DBI} + S_{\rm CS} \, . 
\ee
Let us first discuss the DBI action. Its general form for one of the disconnected D8- and  $\overline{\rm D8}$-branes in the chirally 
symmetric phase is
\be
S_{\rm DBI} = T_8 V_4 \int d^4x \int_{U_T}^\infty dU\, e^{-\Phi}\sqrt{{\rm det}(g+2\pi\alpha'F)} \, , 
\ee
where $\alpha'=\ell_s^2$ with the string length $\ell_s$, the D8-brane tension $T_8=(2\pi)^{-8}\ell_s^{-9}$, the volume of the unit four-sphere
$V_4\equiv 8\pi^2/3$, and the dilaton $e^{\Phi} = g_s(U/R)^{3/4}$ with the string coupling $g_s$. For brevity we have denoted the space-time 
integral by $d^4x$ although
it is actually a Euclidean integral $d\tau d^3x$ with imaginary time $\tau$, such that the integral over a space-time independent 
integrand (which is all we need in our calculation) yields $V/T$ with the three-volume $V$. The DBI action for one half of the connected branes
in the chirally broken phase is given by the same expression with the lower integration boundary $U_T$ replaced by $U_0$.
(In this general section, we shall give the expressions for the symmetric phase, but the broken phase is easily obtained via this 
simple replacement.) 

In our ansatz the only nonzero field strength 
components are $F_{u0}$, $F_{u3}$, $B\equiv F_{12}$. The field strength $F_{12}$ is constant 
in the bulk and corresponds to a homogeneous magnetic field in the spatial 3-direction. Since the gauge symmetry on the flavor branes 
corresponds to a global symmetry at the boundary, $B$ is not a dynamical magnetic field. However, this is not problematic in our context
where we are interested in a fixed background magnetic field. 

We introduce the dimensionless quantities  
\be \label{dimless}
a_{0,3}\equiv \frac{2\pi\alpha'}{R} A_{0,3} \, , \qquad b\equiv 2\pi\alpha' B \, , \qquad u\equiv \frac{U}{R} \, , \qquad x_4\equiv \frac{X_4}{R}
\, , 
\ee
where $A_0$, $A_3$ are the dimensionful gauge fields. Then, with the relation 
\be
R^3 = \pi g_s N_c \ell_s^3  \, , 
\ee
we can write the DBI action in the convenient form, also used in refs.\ \cite{Lifschytz:2009sz,Bergman:2008qv}, 
\be \label{SDBI}
S_{\rm DBI} = \frac{\cal N}{2} \int d^4 x \int_{u_T}^\infty du\,\sqrt{u^5+b^2u^2}\sqrt{1+fa_3'^2-a_0'^2+u^3fx_4'^2} \, .
\ee
Here the prime denotes the derivative with respect to $u$,   
\be
{\cal N}\equiv \frac{N_c}{6\pi^2}\frac{R^2}{(2\pi\alpha')^3}  \, , 
\ee
and 
\be \label{defuT}
f(u) = 1-\frac{u_T^3}{u^3} \, , \qquad u_T = \left(\frac{4\pi}{3}\right)^2 t^2 \, ,
\ee
with the dimensionless temperature 
\be
t\equiv TR \, .
\ee

The CS action is (in the gauge $A_u=0$)
\be
S_{\rm CS} = \frac{N_c}{24\pi^2}\int d^4x\int_{U_T}^\infty dU\,A_\mu F_{u\nu}F_{\rho\sigma}\epsilon^{\mu\nu\rho\sigma} \, ,
\ee
where $\mu,\nu, \ldots = 0,1,2,3$ and $\epsilon_{0123}=+1$. Strictly speaking, this CS action is the action on the left-handed brane (D8-brane). The corresponding
action for the right-handed brane ($\overline{\rm D8}$-brane) has an overall minus sign. To avoid complications in the notation
such as introducing left- and right handed gauge fields we shall only 
write expressions for the left-handed brane. This is sufficient for our 
purpose since we are mainly interested in the free energy of the
system, which does not distinguish between the left- and right-handed fermions. There are of course quantities, such as the currents, where the
sign of the CS action becomes relevant. 

Within the above ansatz and using our dimensionless quantities, the CS part 
can be written as
\be \label{SCS}
S_{\rm CS} = \frac{\cal N}{2}\int d^4x\int_{u_T}^{\infty} du\, \Big[\partial_2 a_1(a_0\partial_ua_3-a_3\partial_ua_0)+
a_1(\partial_ua_0\partial_2a_3-\partial_ua_3\partial_2a_0)\Big] \, .
\ee
Here we have kept all terms that contribute to the equations of motion, although the two terms $\propto a_1$ vanish in our on-shell action
since $a_3$ and $a_0$ do not depend on $x_2$.

If we worked with the action as given by eqs.\ (\ref{SDBI}) and (\ref{SCS}), we would encounter an ambiguity in the currents: in the 
presence of a homogeneous magnetic field the currents defined via the usual AdS/CFT dictionary would deviate from the currents defined through  
thermodynamic relations. This problem was pointed out in refs.\ \cite{Bergman:2008qv,Rebhan:2009vc}. In ref.\ \cite{Bergman:2008qv} 
a modified action was suggested, 
\be
S'= S_{\rm DBI} + S_{\rm CS} + \Delta S \, ,  
\ee
with the additional contribution 
\be
\Delta S = \frac{\cal N}{4}\int d^4x \int_{u_T}^\infty du \, \Big\{ \partial_2\left[a_1\left(a_0\partial_ua_3-a_3 \partial_u a_0\right)\right]-
\partial_u\left[a_1\left(a_0\partial_2 a_3-a_3\partial_2a_0\right)\right]\Big\} \, .
\ee
This term does not change the equations of motion since it is a boundary term, but it is not a usual holographic renormalization
because the first term in the curly brackets is a term at the {\em spatial} boundary, not the holographic boundary. The new action $S'$ 
is invariant under residual gauge transformations which do not vanish at the spatial boundary \cite{Bergman:2008qv} and removes the 
ambiguity in the currents.  However,  
the correspondingly modified currents do not satisfy
correct anomaly equations and reproduce the expected anomalous
conductivities only up to a factor
\cite{Rebhan:2009vc}. Here we are interested in
the phase diagram, and not primarily in the
anomalous conductivities, so we do not attempt to 
resolve this subtle issue; we simply follow the prescription with the modified action $S'$. The on-shell contribution of $\Delta S$ is given solely by the term at the spatial boundary.
This term becomes simply one half of the original CS part, and thus adding $\Delta S$ effectively amounts to 
multiplying the original CS action by 3/2. Therefore, we expect our results 
to differ quantitatively, but not qualitatively, when we use the original action $S$ instead.
We can write our on-shell action for the left-handed flavor brane as
\bea\label{DBICS}
S_{\rm on-shell}' &=& \frac{V{\cal N}}{2T} \int_{u_T}^{\infty} du\,\sqrt{u^5+b^2u^2}\sqrt{1+fa_3'^2-a_0'^2+u^3fx_4'^2} \non[2ex]
&& +\, \frac{3V{\cal N}}{4T}\,  b \int_{u_T}^{\infty} du\, (a_3a_0'-a_0a_3') \, ,
\eea
where we used $\partial_2 a_1=-b$, and where the trivial space-time integral has been performed.
 
The equations of motion for $a_0$, $a_3$, and $x_4$ can be derived from eqs.\ (\ref{SDBI}) and (\ref{SCS}),
\begin{subequations} \label{EOM1}
\bea
\partial_u\left(\frac{a_0'\sqrt{u^5+b^2u^2}}{\sqrt{1+fa_3'^2-a_0'^2+u^3fx_4'^2}}\right) &=& 3ba_3' \, ,\label{EOM1a0}\\
\partial_u\left(\frac{f\,a_3'\sqrt{u^5+b^2u^2}}{\sqrt{1+fa_3'^2-a_0'^2+u^3fx_4'^2}}\right) &=&  3ba_0' \, ,\label{EOM1a3}\\
\partial_u\left(\frac{u^3f\,x_4'\sqrt{u^5+b^2u^2}}{\sqrt{1+fa_3'^2-a_0'^2+u^3fx_4'^2}}\right) &=& 0 \label{EOM1x4}\, .
\eea
\end{subequations}
The left-hand (right-hand) sides of these equations originate from the DBI (CS) part of the action. Since the  
CS part does not depend on $x_4$, the right-hand side of eq.\ (\ref{EOM1x4}) vanishes. 

According to the AdS/CFT dictionary, the current is defined as 
\be
{\cal J}^\mu = - \left.\frac{\partial {\cal L}}{\partial A_\mu'}\right|_{u=\infty} \, .
\ee
Consequently, the 0- and 3-components of the (left-handed) current are 
\begin{subequations} \label{j0j3}
\bea
{\cal J}^0 &=&  \frac{2\pi\alpha'}{R}\frac{\cal N}{2}\left(a_0' u^{5/2}-\frac{3}{2}b\,a_3\right)_{u=\infty}
=\frac{2\pi\alpha'}{R}\frac{\cal N}{2}\left[\frac{3}{2}b\,a_3(\infty)+C\right]
\, , \label{j0}\\
{\cal J}^3 &=&  
-\frac{2\pi\alpha'}{R}\frac{\cal N}{2}\left(a_3' u^{5/2}-\frac{3}{2}b\,a_0\right)_{u=\infty}=
-\frac{2\pi\alpha'}{R}\frac{\cal N}{2}\left[\frac{3}{2}b\,a_0(\infty)+D\right] \, , \label{j3}
\eea
\end{subequations}
where, in the second equality of each line, we have used the integrated form of the equations of motion (\ref{EOM1})
with the integration constants $C$ and 
$D$. Due to our use of $S'$, these results are identical to the ones which are obtained by taking the derivative of the free energy $\Omega$
with respect to the corresponding source. For example, ${\cal J}^0$ is the charge density which is also obtained by the negative of the 
derivative of $\Omega$ with respect to the chemical potential. Since $\Omega$ will turn out to be very complicated in general, 
eq.\ (\ref{j0}) yields a simple alternative way to compute the density.

In the subsequent sections we shall solve the equations of motion. For all $t$, $\mu$, and $b$ there are
two classes of solutions. One with $x_4'=0$, corresponding to straight, disconnected flavor branes and thus the chirally symmetric
phase, and one with $x_4'\neq 0$, corresponding to curved, connected flavor branes and thus the chirally broken phase, see fig.\ \ref{figcylinders}.
The general solution of the equations has to be found numerically, but we shall discuss various limits where semi-analytic solutions
can be found. The solutions will  
then be inserted into the action in order to compare the free energies of the 
chirally broken and symmetric phase. This will lead us to our main result, the chiral phase transition as a critical surface in 
the $t$-$\mu$-$b$ parameter space.

\section{Chirally broken phase}
\label{sec:broken}

\subsection{Solution in the $f\simeq 1$ approximation}
\label{sec:general}

In general, the case of connected flavor branes is the more complicated one since besides the 
gauge fields $a_0$ and $a_3$ the equations of motion also contain the nontrivial function $x_4$.
We simplify this case by approximating
\be \label{f1}
f(u) \simeq 1 
\ee
for all $u$ on the flavor branes. This approximation is valid for sufficiently large $u_0\gg u_T$ since for all $u$ on the 
flavor branes we have $u\ge u_0$, see fig.\ \ref{figcylinders}. In principle,
the approximation can -- at a fixed temperature and thus fixed $u_T$ -- be made arbitrarily good by decreasing the 
asymptotic distance of the flavor branes $L$. However, we have to keep in mind that we are interested in the critical temperature 
for the chiral phase transition. Suppose we choose $L$ very small such that $f\simeq 1$ is a good approximation at some
small temperature. Then, increasing the temperature and keeping $L$ fixed tends to invalidate our approximation because $u_T\propto T^2$
increases and approaches $u_0$. But at some 
critical temperature the chirally symmetric phase takes over and thus our approximation only needs to be valid at temperatures below
this (a priori unknown) critical temperature. At $b=\mu=0$, where the full treatment is simple, we have checked that our result for the 
critical temperature deviates by about 10\% from the full result, see sec.\ \ref{sec:B0}.
In the general case we have only solved the equations of motion in the 
limit $f\simeq 1$, and thus have no quantitative comparison with the full result, but 
we have checked that the transition takes over before severe artifacts such as $u_0<u_T$ occur in our approximation. 
Note also that within the approximation $f\simeq 1$ the broken phase becomes independent of $T$. (The chiral phase transition 
will still depend on $T$ due to the $T$-dependence of the chirally restored phase.)

With eq.\ (\ref{f1}) the integrated version of the equations of motion becomes
\begin{subequations} \label{EOM2}
\bea
\frac{a_0'\sqrt{u^5+b^2u^2}}{\sqrt{1+a_3'^2-a_0'^2+u^3x_4'^2}} &=& 3ba_3 + c \, ,\label{EOM2a0}\\[1.5ex]
\frac{\,a_3'\sqrt{u^5+b^2u^2}}{\sqrt{1+a_3'^2-a_0'^2+u^3x_4'^2}} &=& 3ba_0 + d \, ,\label{EOM2a3}\\[1.5ex]
\frac{u^3\,x_4'\sqrt{u^5+b^2u^2}}{\sqrt{1+a_3'^2-a_0'^2+u^3x_4'^2}} &=& k \label{EOM2x4}\, ,
\eea
\end{subequations}
with integration constants $c$, $d$, and $k$. Our boundary conditions are 
\begin{subequations} \label{boundarybroken}
\bea
a_0(\infty) &=& \mu  \, , \qquad \frac{a_0'(u_0)}{a_3'(u_0)} = 0 \, , \\[1.5ex]
a_3(\infty) &=& \jmath \, , \qquad a_3(u_0) = 0 \, , \label{bounda3}\\[1.5ex]
x_4'(u_0) &=& \infty \, , \qquad \frac{\ell}{2}=\int_{u_0}^{\infty}du\,x_4' \, , \label{boundx4}
\eea
\end{subequations}
where we have introduced the dimensionless separation 
\be
\ell \equiv \frac{L}{R} \, .
\ee
These boundary conditions arise as follows. First, we require the temporal component of the gauge field to approach the quark chemical potential 
$\mu$ at the holographic boundary. Since the chemical potential is the same for left- and right-handed quarks, $a_0$ must be symmetric, i.e., 
if $a_0$ is a smooth function along the entire connected branes, its derivative at the tip of the brane must vanish, $a_0'(u_0)=0$.
It turns out, however, that in the given choice of coordinates the more general version of this boundary condition 
is $a_0'(u_0)/a_3'(u_0)=0.$ This takes into account that for finite magnetic field $a_0$ has a cusp at $u=u_0$, i.e., $a_0'(u\to u_0)$ 
approaches two values with the same magnitude but opposite sign depending on whether one approaches $u_0$ from the left or right.  
At the same time $a_3'(u_0)$ becomes infinite. This apparent singularity can be removed by a coordinate change, for instance to the 
variable $z$ used in ref.\ \cite{Rebhan:2009vc}, defined through $u=(u_0^3+u_0z^2)^{1/3}$ (and vice versa, i.e., the smooth solutions in 
terms of $z$ of ref.\ \cite{Rebhan:2009vc} acquire the same cusp in $a_0$ after changing the coordinate to $u$). 
 
For the spatial component in the direction of the magnetic field $a_3$ we must 
allow for a nonzero value $\jmath$ at the holographic boundary which has to be determined dynamically by minimization of the free energy. It 
has been shown for the technically simpler cases of maximally separated branes and/or the Yang-Mills 
approximation of the DBI action \cite{Thompson:2008qw,Bergman:2008qv,Rebhan:2008ur} that $\jmath$ assumes a nonzero value in the presence 
of a magnetic field. This corresponds to an anisotropic chiral condensate and thus, viewing this condensate as a superfluid, 
$\jmath$ corresponds to a supercurrent \cite{Rebhan:2008ur}. As a consequence, the system acquires nonzero baryon number, even for baryon
chemical potentials smaller than the baryon mass \cite{Thompson:2008qw,Son:2007ny}. 
Since the chiral condensate 
carries axial, not vector, charge, $\jmath$ is an axial supercurrent. Consequently, the boundary value at the other asymptotic end of the 
connected branes (where the right-handed fermions live) must be $-\jmath$, leading to an antisymmetric
gauge field $a_3$ and thus to the boundary condition $a_3(u_0) = 0$. 

Finally, the first boundary condition in eq.\ (\ref{boundx4})
says that the connected branes ``turn around'' smoothly at $u=u_0$ while the second one says that the asymptotic (dimensionless) separation
of the branes is $\ell$.

We can solve eqs.\ (\ref{EOM2}) semi-analytically by generalizing the method introduced in 
ref.\ \cite{Bergman:2008qv} to the case $x_4'\neq 0$. In this way, the differential equations can be reduced to 
two coupled algebraic equations which have to be solved numerically. We defer the details of this procedure to 
appendix \ref{app:broken}. The result can be written as follows. We introduce the constant $\eta$ via
\be \label{eta}
u_0^{3/2}\eta = \lim_{u\to u_0}\frac{a_3'(u)}{x_4'(u)}  \, . 
\ee
A nonzero $\eta$ implies that not only $x_4'$, but also the derivative of $a_3$ becomes infinite at the tip of the branes $u_0$. 
Moreover, we define the new variable $y$ through
\be \label{yu}
y(u) = 3b\sqrt{1+\eta^2}\int_{u_0}^u \frac{v^{3/2}dv}{\sqrt{g(v)}} \, ,
\ee
with 
\be\label{defg}
g(u)\equiv (\eta^2+1)(u^8+b^2u^5)-\left(\eta^2\frac{u^3}{u_0^3}+1\right)(u_0^8+b^2u_0^5) \, .
\ee
In terms of these quantities, the solution for the gauge fields is
\begin{subequations} \label{ay}
\bea
a_0(y) &=& \mu + \frac{\jmath}{\sinh y_\infty} (\cosh y -\cosh y_\infty)  \, ,\label{ay1} \\
a_3(y) &=& \frac{\jmath}{\sinh y_\infty}\sinh y  \, , \label{ay2} 
\eea
\end{subequations}  
where $y_\infty \equiv  y(u=\infty)$, and the embedding of one half of the connected flavor branes is given by
\be \label{x4u}
x_4(u) = u_0^{3/2}\sqrt{u_0^5+b^2u_0^2}\int_{u_0}^u\frac{dv}{v^{3/2}\sqrt{g(v)}} \, . 
\ee
The functions $a_0$, $a_3$, $x_4$ are written in terms of $\mu$, $b$ (which are the externally fixed physical parameters), 
$\jmath$ (which has to be determined from minimizing the free energy), and the constants $u_0$, $\eta$, which are functions
of $\ell$, $\jmath$, and $b$ and given by the coupled equations
\begin{subequations} \label{u0eta}
\bea
\frac{\ell}{2} &=& u_0^{3/2}\sqrt{u_0^5+b^2u_0^2}\int_{u_0}^\infty\frac{du}{u^{3/2}\sqrt{g(u)}} \, , \label{u0eta1}\\[1.5ex]
\frac{\jmath}{\sinh y_\infty} &=& \frac{\sqrt{u_0^5+b^2u_0^2}}{3b}\frac{\eta}{\sqrt{1+\eta^2}} \, . \label{u0eta2}
\eea
\end{subequations}
The dependence on the separation $\ell$ can be eliminated by rescaling
\be \label{rescale}
b\to \ell^3 b \, ,  \qquad \jmath\to \ell^2 \jmath \, , \qquad u_0\to \ell^2 u_0 \, .
\ee
($\eta$ and $y_\infty$ are invariant under rescaling with $\ell$.)
Employing these rescalings and changing the integration variable $u\to \ell^2 u$ is equivalent to simply setting $\ell=1$ in 
eq.\ (\ref{u0eta1}). 

We could now proceed by solving eqs.\ (\ref{u0eta}) for all $\mu$, $b$, and $\jmath$, insert the result into the 
solutions (\ref{ay}) and (\ref{x4u}), these solutions into the on-shell action (\ref{DBICS}) and minimize the resulting free energy 
with respect to $\jmath$. However, there is a simpler way to determine the supercurrent $\jmath$. We recall that the total axial current 
${\cal J}^3$ is obtained by taking the derivative of the free energy with respect to the corresponding source. Here, $\jmath$ plays the role
of that source and thus we conclude that $\jmath$ extremizes the free energy if ${\cal J}^3 = 0$,
which implies, using eq.\ (\ref{j3}),
\be \label{jmin}
\jmath = \frac{\mu}{2}\tanh y_\infty \, .
\ee
This result can now be inserted into eq.\ (\ref{u0eta2}) which eliminates $\jmath$ from the numerical calculation. 
Written in this way, $\jmath$ is the same as in ref.\ \cite{Bergman:2008qv}, but note that $y_\infty$ is different in this 
reference. The reason is that there maximally separated flavor 
branes were considered, and thus the chirally broken phase was discussed in the confined geometry. 

The free energy is
\be
\Omega = 2\frac{T}{V} S'_{\rm on-shell} \, ,
\ee
where the factor 2 takes into account both halves of the connected branes and where $S'_{\rm on-shell}$ is given in eq.\ (\ref{DBICS}).
After inserting the solutions of the equations of motion and after some algebra we can write the free energy as
\bea \label{OmegaB}
\Omega_\cup &=& {\cal N}\int_{u_0}^\infty du\,\frac{u^{3/2}}{\sqrt{\eta^2+1}}\frac{(\eta^2+1)(u^5+b^2u^2)-\frac{\eta^2}{2}(u_0^5+b^2u_0^2)}
{\sqrt{g(u)}} \non[1.5ex]
&&
+\,\frac{3}{2}{\cal N}b\,\underbrace{\left(\sqrt{\jmath^2+\frac{u_0^5+b^2u_0^2}{(3b)^2} 
\frac{\eta^2}{1+\eta^2}}-\mu\right)\jmath}_{\displaystyle{-\frac{\mu^2}{4}\tanh y_\infty}} \, , 
\eea
where the result below the curly bracket eliminates $\jmath$ and has been obtained by using eqs.\ (\ref{u0eta2}) and (\ref{jmin}).
Inserting the rescaled quantities from eq.\ (\ref{rescale}) into $\Omega_\cup$ shows that the free energy and the chemical potential
scale as 
\be \label{rescale2}
\Omega \to \ell^7\Omega \, , \qquad \mu\to \ell^2 \mu \, .
\ee
In the remainder of the paper we shall set $\ell=1$ in all plots (except for fig.\ \ref{figsplit}) for convenience. 
The $\ell$ dependence of all curves can easily be recovered with the rescalings (\ref{rescale}) and (\ref{rescale2}).

Finally, we can compute the quark number density $n$. With eq.\ (\ref{j0}) we find 
\be \label{nXb}
n \equiv 2{\cal J}^0 = \frac{3}{2}\frac{2\pi\alpha'}{R}{\cal N}b\jmath = \frac{3}{4}\frac{2\pi\alpha'}{R}{\cal N} b \mu\tanh y_\infty \, .
\ee
(Recall that eqs.\ (\ref{j0j3}) only take into account one half of the branes, hence the factor 2 in the definition of the total 
density $n$.)

\subsection{Discontinuity in the density}
\label{sec:limit}

\FIGURE[t]{\label{figOmegaj}
{ \centerline{\def\epsfsize#1#2{0.72#1}
\epsfbox{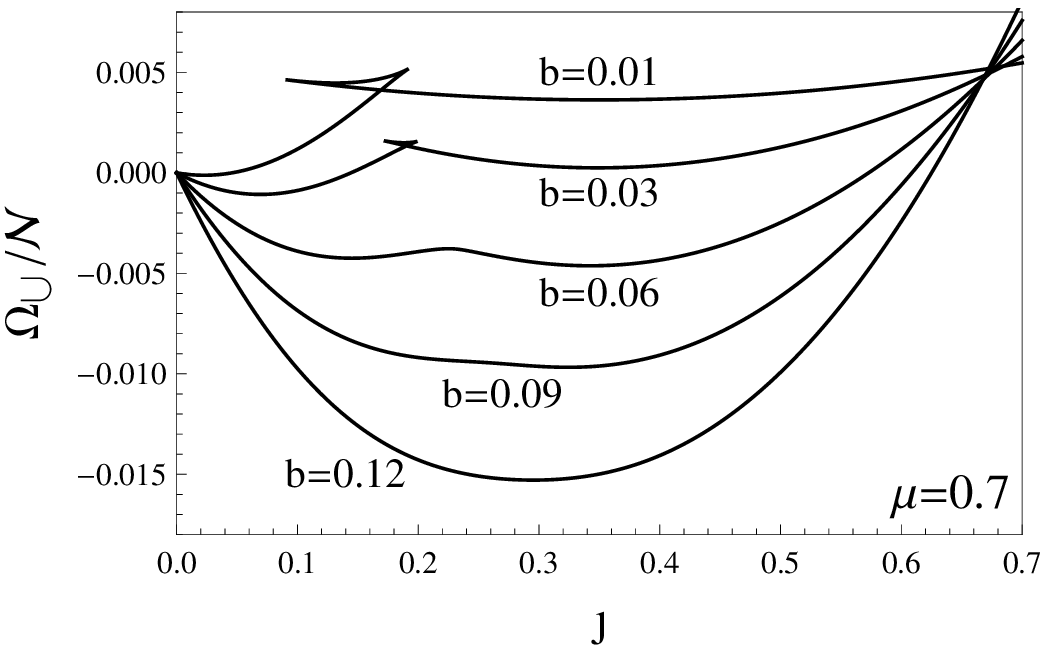}
\def\epsfsize#1#2{0.69#1}
\epsfbox{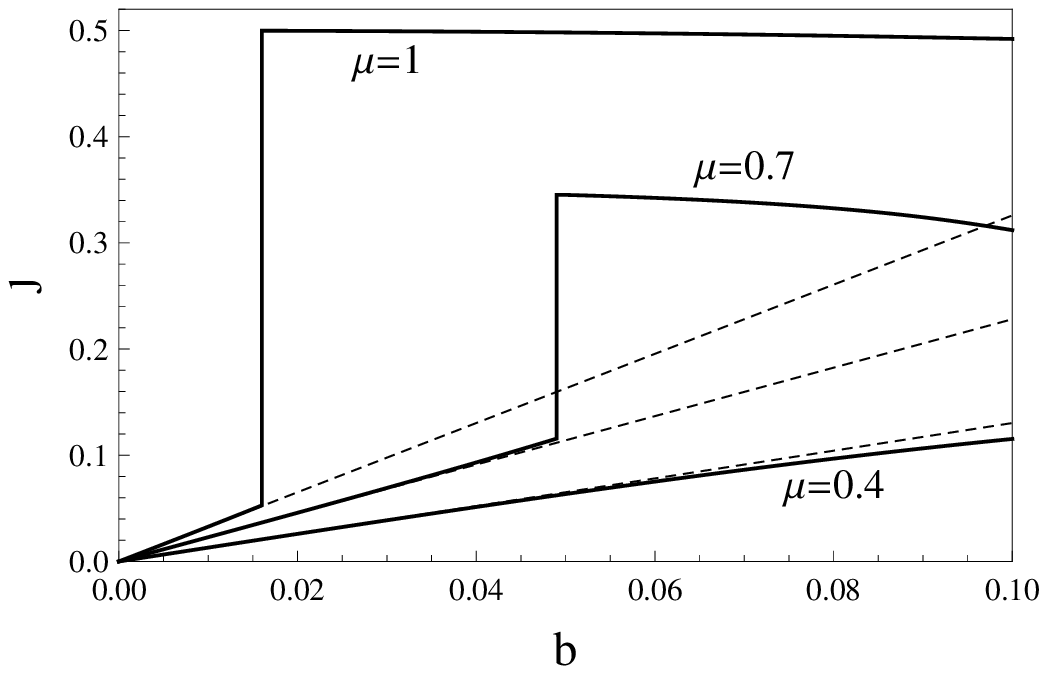}}
}
\caption{Left panel: free energy $\Omega_\cup$ as a function of the supercurrent $\jmath$ in the chirally broken phase for a fixed chemical 
potential
$\mu$ and several values of the magnetic field $b$ [in the dimensionless units given in eq.\ (\ref{dimless}), and with $\ell=1$ which 
is equivalent to using the rescaled quantities from eqs.\ (\ref{rescale}) and (\ref{rescale2})]. 
Since $\jmath$ is dynamically determined from minimization of $\Omega_\cup$, we 
see that for the given $\mu$ there is a first-order phase transition where $\jmath$ is discontinuous. 
Right panel: supercurrent $\jmath$ as a function of $b$ for three values of $\mu$. The solid lines are the full numerical
result while the dashed lines are the linear approximations for small $b$ according to eq.\ (\ref{jlin}). The middle curve $\mu=0.7$ corresponds 
to the potentials shown in the left panel. All three curves approach the asymptotic limit $\jmath\to\mu/2$ for large magnetic field, i.e., 
the upper two curves change their curvature to approach this limit on a $b$ scale much larger than shown here.
 }  
}

\FIGURE[t]{\label{figjjump}
{ \centerline{\def\epsfsize#1#2{0.8#1}
\epsfbox{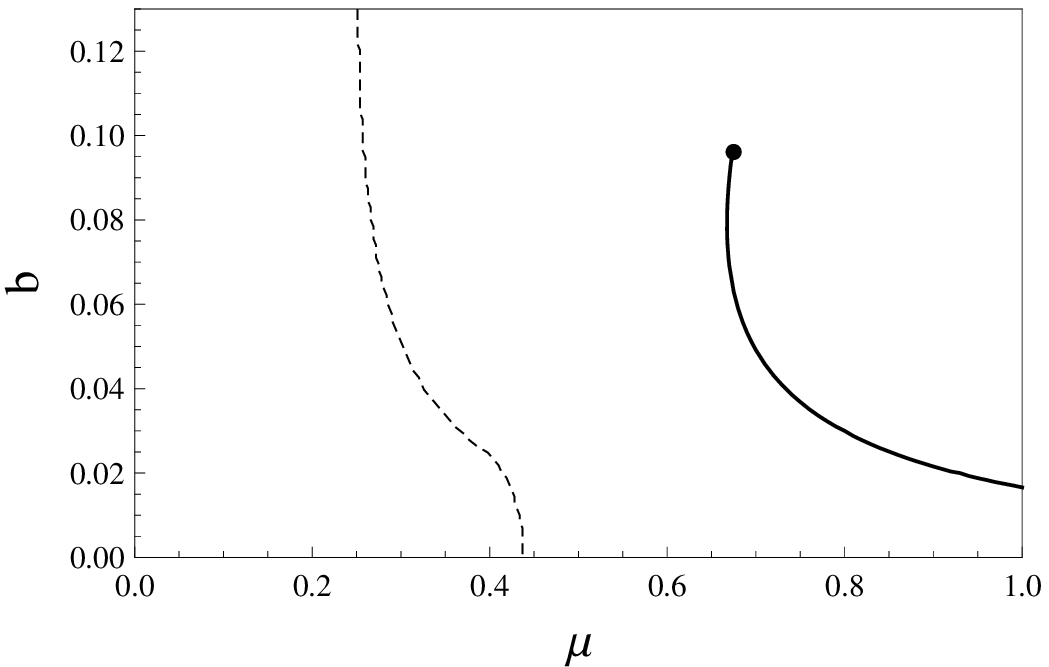}
}
}
\caption{First-order phase transition (solid line) in the $\mu$-$b$ plane for the chirally broken phase with $b$, $\mu$ in the 
same units as in fig.\ \ref{figOmegaj}. The supercurrent
(and thus the baryon density) is discontinuous across this line, as shown in fig.\ \ref{figOmegaj}. The dashed line
is the chiral phase transition at $T=0$ (see sec.\ \ref{sec:transition}). On the right-hand side of the dashed line, the ground state turns out 
to be chirally symmetric (disconnected flavor branes) such that the solid line is only relevant for a metastable state.
 }  
}

Although we can compute the supercurrent $\jmath$ directly from eq.\ (\ref{jmin}), let us first discuss the form of the free energy 
as a function of $\jmath$. Solving eqs.\ (\ref{u0eta}) shows that there are parameter regions where there is a unique solution for the 
pair $(u_0,\eta)$ and parameter regions where there are three solutions. This is reflected in the free energy shown in the left 
panel of fig.\ \ref{figOmegaj}. To obtain this plot we have renormalized $\Omega$ by subtracting the vacuum contribution
\be \label{vacXb}
\Omega_\cup(\mu=\jmath=0) = {\cal N}\int_{u_0}^\infty du\,u^{3/2}\frac{u^5+b^2u^2}{\sqrt{u^8+b^2u^5-(u_0^8+b^2u_0^5)}} \, , 
\ee
where we have used that $\eta=0$ for $\jmath=0$, see eq.\ (\ref{u0eta2}). The curves of the renormalized potential as a function 
of $b$ show that there is a first-order phase transition where $\jmath$ is discontinuous and thus, due to eq.\ (\ref{nXb}), 
also the baryon density $n/N_c$.   The discontinuity is shown explicitly
in the right panel where $\jmath$ is plotted as a function of $b$ for three different values of $\mu$. The full result has been obtained 
numerically, but we can easily find analytic approximations for small and large values for the magnetic field. For small magnetic fields
and small $\jmath$, eqs.\ (\ref{u0eta}) give
\be \label{u0smallb}
u_0\simeq \left(\frac{2P_1}{\ell}\right)^2 \,, \qquad \eta \simeq \frac{\jmath}{u_0P_2} \, , 
\ee
with 
\be \label{P1P2}
P_1\equiv \int_1^\infty \frac{du}{u^{3/2}\sqrt{u^8-1}}=\frac{2\sqrt{\pi}\,\Gamma\left(\frac{9}{16}\right)}{\Gamma\left(\frac{1}{16}\right)} \, , 
\qquad P_2\equiv \int_1^\infty \frac{u^{3/2}\,du}{\sqrt{u^8-1}}=\frac{\sqrt{\pi}\,
\Gamma\left(\frac{3}{16}\right)}{8\Gamma\left(\frac{11}{16}\right)} \, .
\ee
We can thus approximate $\tanh y_\infty \simeq y_\infty$ and eq.\ (\ref{jmin}) becomes
\be \label{jlin}
\jmath \simeq \frac{3P_2\ell^3}{16 P_1^3}\,\mu \,b   \qquad (\mbox{small}\; b) \, .
\ee
This simple linear form for $\jmath$ is compared to the full result in the right panel of fig.\ \ref{figOmegaj}. For large 
magnetic fields, $\tanh y_\infty \simeq 1$ and thus $\jmath$ approaches $\mu/2$.

In fig.\ \ref{figjjump} we show the discontinuity in the baryon density in the $b$-$\mu$ plane. As suggested from the right panel of 
fig.\ \ref{figOmegaj}, the discontinuity is only present for sufficiently large chemical potentials. The first-order
phase transition line terminates in a critical point and approaches the $\mu$ axis for small magnetic fields. 
The figure also shows the chiral phase transition at $T=0$, to be computed and discussed in sec.\ \ref{sec:transition}.
On the right-hand side of this line, chiral symmetry is restored. Therefore, the discontinuity in the density 
only occurs in a metastable phase and is probably of little physical relevance. We shall thus not display it in the 
phase diagrams in the subsequent sections.

\section{Chirally symmetric phase}
\label{sec:symmetric}

The chirally symmetric phase has been considered in ref.\ \cite{Lifschytz:2009sz} within the same setup as discussed here. 
Nevertheless we shall
discuss some of the details of this phase before we come to the chiral phase transition. 
One reason is that we work at fixed chemical potential, while in ref.\ \cite{Lifschytz:2009sz}
the density was held fixed. Furthermore, we shall elaborate on a discontinuity in the charge density within this phase, which 
resembles a transition to the lowest Landau level. A physical understanding of this discontinuity will turn out to be useful
in the comparison of our phase diagrams with NJL model calculations. 

In the case of disconnected flavor branes the integrated form of the equations of motion (\ref{EOM1}) becomes
\begin{subequations} \label{EOMXs}
\bea
\frac{a_0'\sqrt{u^5+b^2u^2}}{\sqrt{1+fa_3'^2-a_0'^2}} &=& 3ba_3 + C \, , \label{EOMXs1}\\
\frac{f\,a_3'\sqrt{u^5+b^2u^2}}{\sqrt{1+fa_3'^2-a_0'^2}} &=& 3ba_0 + D \, , \label{EOMXs2}
\eea
\end{subequations}
with integration constants $C$ and $D$. 
Since the branes are straight, we have set $x_4'=0$. In this sense, the equations are simpler than for the 
case of connected branes. However, now we cannot use the approximation $f\simeq 1$ because the branes extend all the way down to  
$u=u_T$, see fig.\ \ref{figcylinders}. In this sense, the equations are more difficult than the ones for the connected branes.
In general, we have to solve these equations numerically. Our boundary conditions are
\be \label{boundaryXs}
a_0(\infty) = \mu \, , \qquad a_0(u_T)=a_3(\infty) = 0 \, .
\ee
As in the chirally broken phase, the value of $a_0$ at the holographic boundary is identified with the chemical potential. In contrast to the 
broken phase, the boundary value of $a_3$ vanishes because there are no Goldstone modes without spontaneous symmetry breaking, 
and thus there cannot be any supercurrent of these modes. We also require $a_0$ to vanish at the horizon,
which is a regularity constraint\footnote{In Ref.~\cite{Gynther:2010ed} it was
argued that this regularity constraint needs to be abandoned in the
case of an axial chemical potential. However, for an ordinary
chemical potential as considered here, gauge invariance implies that this constraint
is not a physical restriction.}
 \cite{Horigome:2006xu}. For $a_3$, there is a priori no condition at the horizon. Because of $f(u_T)=a_0(u_T)=0$,
eq.\ (\ref{EOMXs2}) immediately yields 
\be \label{D0}
D=0 \, .
\ee
(Provided that $a_3'(u_T)$ is finite, which is true in all solutions we consider.) The numerical evaluation of eqs.\ (\ref{EOMXs}) can be done 
with the ``shooting method'': we consider the two differential equations
as an initial value problem by imposing the initial values at the boundary $u=\infty$ according to eq.\ (\ref{boundaryXs}). Then we solve
the equations by letting the gauge fields evolve from $u=\infty$ to $u=u_T$ for all $C$ from an appropriately chosen interval. 
(It turned out to be useful to implement this procedure by promoting the ordinary differential equations to partial differential equations 
with the additional variable $C$.) Then we determine
the value(s) of $C$ for which the gauge field $a_0$ is ``shot'' to its correct value at the horizon, $a_0(u_T)=0$. (Since in some cases two of these 
values for $C$ are very close to each other, it is more convenient to reparametrize $C\to 3b\mu\coth z_\infty$ in the numerics, motivated by the 
zero-temperature solution, see below.)

\subsection{Zero-temperature limit and "Landau level" transition}
\label{sec:Tzero2}

For $T=0$, we have $u_T=0$ and thus $f=1$. In this case the equations (\ref{EOMXs}) can be solved semi-analytically. The solution is
(see appendix \ref{app:symmetric} for details)
\begin{subequations} \label{solXs}
\bea
a_0(z) &=& \frac{\mu}{\sinh z_\infty} \sinh z \, , \\
a_3(z) &=& \frac{\mu}{\sinh z_\infty} (\cosh z - \cosh z_\infty)  \, ,
\eea
\end{subequations}
with the new variable 
\be \label{zu}
z(u) = 3b\int_0^u\frac{dv}{\sqrt{v^5+b^2v^2+\frac{(3b\mu)^2}{\sinh^2 z_\infty}}} \, ,
\ee
where $z_\infty\equiv z(u=\infty)$ has to be determined numerically from the relation
\be \label{z8}
z_\infty = 3b\int_0^\infty\frac{du}{\sqrt{u^5+b^2u^2+\frac{(3b\mu)^2}{\sinh^2 z_\infty}}} \, .
\ee 
Inserting the solution (\ref{solXs}) into the on-shell action (\ref{DBICS}) yields the free energy 
\be \label{OmegaXs}
\Omega_{||}(t=0) = {\cal N}\int_0^\infty 
du\,\frac{u^5+b^2u^2+\frac{1}{2}\frac{(3b\mu)^2}{\sinh^2 z_\infty}}{\sqrt{u^5+b^2u^2+\frac{(3b\mu)^2}{\sinh^2 z_\infty}}}
-\frac{3}{2}{\cal N}b\mu^2\coth z_\infty \, .
\ee
The quark number density is obtained from eq.\ (\ref{j0}),
\be
n \equiv 2{\cal J}^0 = 3\frac{2\pi\alpha'}{R} {\cal N}b\mu\coth z_\infty \, .
\ee
One solution of eq.\ (\ref{z8}) is $z_\infty = \infty$. In this case, the density becomes 
\be \label{nLLL}
n  = \frac{N_c}{2\pi^2} B \mu_q  \, , 
\ee
where $\mu_q$ is the dimensionful quark chemical potential, $\mu_q = R/(2\pi\alpha')\mu$, see eq.\ (\ref{dimless}) for the 
corresponding relation for the gauge fields. 

The numerical calculation shows that in certain regions of the $b$, $\mu$ 
parameter space there are two additional nontrivial solutions for $z_\infty$. Also for nonzero temperatures, where we solve the 
differential equations purely numerically, one or three solutions are found.
\FIGURE[t]{\label{figLLL1}
{ \centerline{\def\epsfsize#1#2{1#1}
\epsfbox{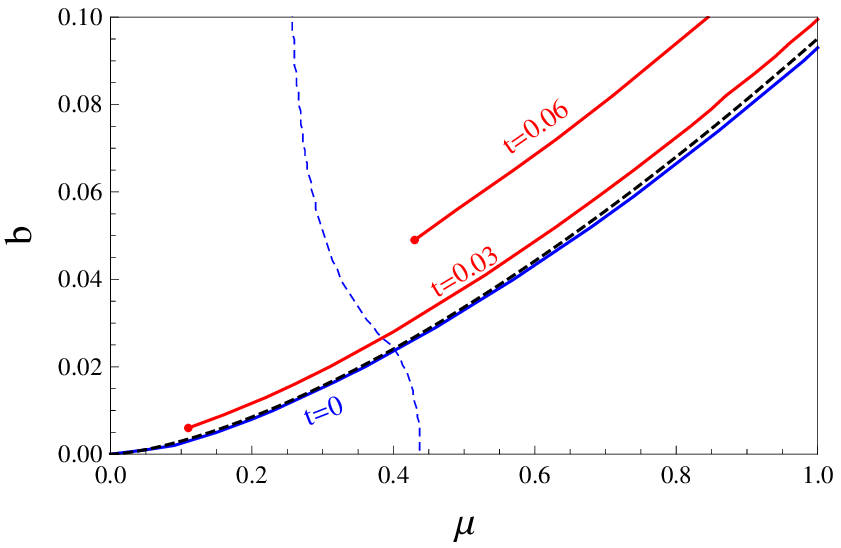}
}
}
\caption{Critical lines in the $b$-$\mu$ plane across which there is a discontinuity in the density of the chirally symmetric
phase for three 
different (dimensionless) temperatures $t$ (to recover the $\ell$ dependence, $t$ has to be replaced by $t\ell$). 
For all nonzero temperatures the critical line ends at a critical point.  We discuss in the text and with the help of figs.\ \ref{figLLL2} 
and \ref{figLLL3} that the critical line is reminiscent
of a Landau level transition, i.e., a population solely in the lowest Landau level above the line and populated higher Landau levels 
below the line (no further transitions between the higher Landau levels are seen in our model). The thin (blue) dashed line is the same line as in 
fig.\ \ref{figjjump}
and indicates the chiral phase transition at $t=0$ (to be computed in sec.\ \ref{sec:transition}), i.e., on the left-hand side of this line  
the (blue) $t=0$ critical line has no physical meaning. The thick (black) dashed line is the analytic approximation (\ref{bcapprox}) 
to the numerical $t=0$ result.      
 }  
}
When we find three solutions $0<z_\infty^{(1)}<z_\infty^{(2)}<z_\infty^{(3)}$, where $z_\infty^{(3)}=\infty$ for $T=0$, the intermediate solution
is never a global minimum of the free energy, while
$z_\infty^{(1)}$ and $z_\infty^{(3)}$ compete for the lowest free energy. 
Where the global minimum jumps from $z_\infty^{(1)}$ to $z_\infty^{(3)}$,
a first-order critical surface appears in the $t$-$\mu$-$b$ parameter space. This surface is 
bounded by a critical line such that two-dimensional cuts through this parameter space, say at fixed temperature, show a critical line
which, for nonzero temperatures, ends at a critical point. This is shown in fig.\ \ref{figLLL1}. For zero temperature, the critical line
is given by the approximate critical magnetic field 
\be \label{bcapprox}
b_c(t=0) \simeq 0.095 \, \mu^{3/2} \, .
\ee
This result is derived in appendix \ref{app:bc} and compared to the full solution in fig.\ \ref{figLLL1}.
The ground state above this critical line is given by the solution $z_\infty=\infty$ and thus the corresponding density by eq.\ (\ref{nLLL}). 
Below the critical line the state with a nontrivial solution  $z_\infty <\infty$ (which depends on $b$ and $\mu$) has the lowest free energy. 
In this case, the density is more complicated. Only for $b\ll b_c(t=0)$, we find the approximate behavior $n \propto \mu^{5/2}$, 
because in this limit $z_\infty\propto b/\mu^{3/2}$, see appendix \ref{app:bc}. For nonzero temperatures, all solutions for 
$z_\infty$ are finite and they continuously merge into each other for sufficiently small $\mu$.

\FIGURE[t]{\label{figLLL2}
{ \centerline{\def\epsfsize#1#2{0.98#1}
\epsfbox{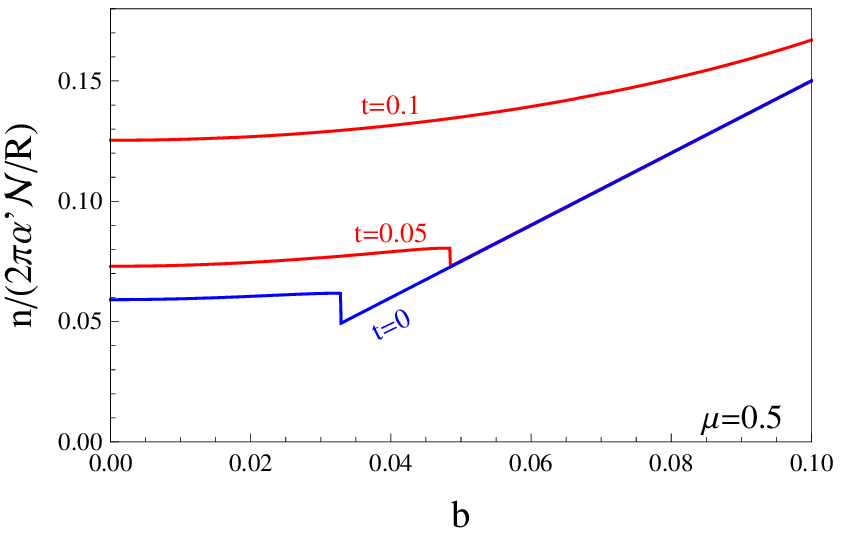}
}
}
\vspace*{-3mm}
\caption{Quark number density $n$ as a function of the magnetic field $b$ for several temperatures $t$ at a fixed chemical 
potential $\mu=0.5$. There is a first-order phase transition at a critical magnetic field
for sufficiently small temperatures. At $t=0$, the density for magnetic fields above this critical value 
is exactly that of a non-interacting Fermi gas in a magnetic field, see eq.\ (\ref{nLLL}). 
 }  
}
\FIGURE[ht]{\label{figLLL3}
{ \centerline{\def\epsfsize#1#2{0.8#1}
\epsfbox{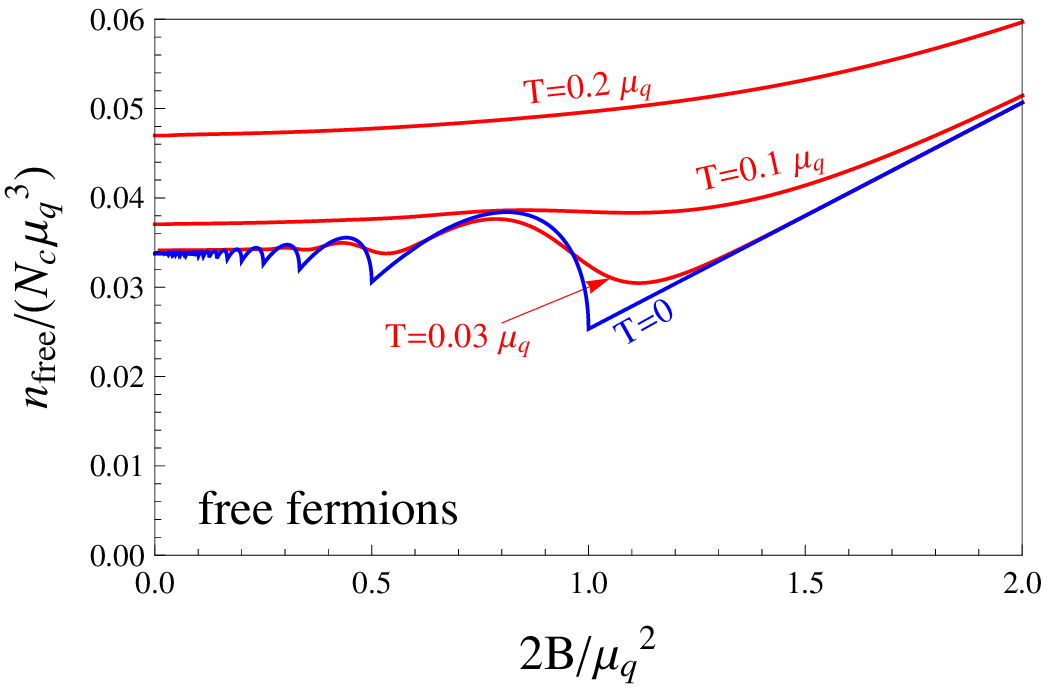}}
}
\vspace*{-3mm}
\caption{Number density for a system of non-interacting, massless fermions as a function of the magnetic field for 
several temperatures (all in appropriate units of the quark chemical potential $\mu_q$). The oscillations are caused by the successive
population of Landau levels and are smeared out for large temperatures. This plot should be compared with the holographic result in 
fig.\ \ref{figLLL2}, which shows some features comparable to a Landau level structure.   
 }  
}

In fig.\ \ref{figLLL2} we show the density as a function of $b$ for several temperatures at a fixed $\mu$. 
There are interesting parallels and differences to the case of free massless fermions in a magnetic field. 
The free energy of $N_c$ non-interacting spin-$\frac{1}{2}$
fermion species 
of charge 1 in a homogeneous magnetic field ${\bf B}=B {\bf e}_3$ is 
\be \label{Omegafree}
\Omega_{\rm free} = -\frac{N_cB}{\pi}\sum_{l=0}^\infty (2-\delta_{l 0}) \sum_{s=\pm} T\int_0^\infty  \frac{dk_3}{2\pi} 
\ln\left[1+e^{-(\epsilon_{k_3,l}-s\mu_q)/T}\right] \, ,
\ee
where $l$ labels the Landau levels. (In general, $B$ has to be replaced by $|qB|$ in this expression, where $q$ is the charge of the fermions.)
The factor $2-\delta_{l 0}$ takes into account that the 
lowest Landau level (LLL) is occupied by a single spin degree of freedom, while all other Landau levels are degenerate with respect to 
both spin projections. The single-particle excitations are  $\epsilon_{k_3,l} = \sqrt{k_3^2+2Bl}$, where $k_3$ is the projection of the 
momentum on the direction of the magnetic field.  The density follows immediately by taking the derivative with respect to $\mu_q$,
\be
n_{\rm free} = \frac{N_cB}{\pi}\sum_{l=0}^\infty (2-\delta_{l 0}) \sum_{s=\pm} s \int_0^\infty  \frac{dk_3}{2\pi} f_F(\epsilon_{k_3,l}-s\mu_q)\,,
\ee
where $f_F(x)\equiv (e^{x/T}+1)^{-1}$ is the Fermi distribution function. At $T=0$, the distribution acquires a sharp Fermi surface and the 
density can be written as 
\be
n_{\rm free}(T=0) = \frac{N_c}{2\pi^2}B\mu_q +\frac{N_cB}{\pi^2}\sum_{l=1}^{\left\lfloor \frac{\mu_q^2}{2B}\right\rfloor}
\sqrt{\mu_q^2-2Bl} \, . \label{nfree}
\ee
Here we have separated the contribution from the LLL which is populated for arbitrarily large $B$.  The 
higher Landau levels $l >0$ are, for a given chemical potential, only populated for sufficiently small magnetic fields which is 
reflected in the upper limit of the sum over $l$. 

We plot $n_{\rm free}$ as a function of the magnetic field in fig.\ \ref{figLLL3}. At $T=0$, there are cusps in the density curve
(i.e., discontinuities in the second derivative of the thermodynamic potential) which are caused by the Landau levels. Coming from large $B$, where
only the LLL is occupied, contributions from higher Landau levels set in successively at each of these cusps. At small $B$, the sum
over discrete levels can be approximated by an integral, and the result approaches the constant $n_{\rm free}(T\!=\!B\!=\!0)= N_c\mu_q^3/(3\pi^2)$
plus a highly oscillatory contribution with amplitude proportional
to $B^{3/2}$.
For arbitrarily small nonzero temperature the cusps are smeared out. The oscillatory behavior survives for small $T$ and then
completely disappears for large $T$. This is due to the smearing of the Fermi surface, i.e., at any nonzero $T$ strictly speaking
all Landau levels are occupied.

We can summarize the comparison of our holographic result for the chirally symmetric phase to the particle picture as follows.
\begin{itemize}

\item {\it Zero temperature.--}
For large magnetic fields, the holographic density behaves {\it exactly} (i.e., all geometric constants of the model drop out) 
like that of a system of non-interacting 
fermions; this can be seen by comparing eqs.\ (\ref{nLLL}) and (\ref{nfree}). In the particle picture, all fermions sit in the LLL
in this limit.

At a certain value of the magnetic field, namely $B=\mu_q^2/2$, the non-interacting system starts to populate the 
first Landau level. This manifests itself in a cusp in the density curve 
corresponding to a second order transition, with infinitely many more
as $B$ is lowered.
In the holographic system there is instead a single first-order phase transition
at the point where, coming from large $B$, the apparent LLL behavior ends. The critical value of $B$ at which this transition happens cannot
be directly compared to the one in the particle picture since it involves the geometric constants of the model such as the curvature radius $R$.   
In dimensionless quantities, this value is $b\simeq 0.0951\mu^{3/2}$, i.e., it goes with a different power of $\mu$ 
than in the case of free fermions. In other words, the effective mass through the magnetic field seems to behave as $B^{2/3}$, not as $B^{1/2}$. 

At small magnetic fields, the density in both systems becomes approximately constant in $B$, for free particles $n_{\rm free}\propto \mu_q^3$, while 
in the Sakai-Sugimoto model $n\propto \mu^{5/2}$. Whereas the free fermion system shows an oscillatory behavior
due to the Landau levels, the holographic result does not seem to know about Landau levels other than $l=0$. 

\item {\it Nonzero temperature.--}
While the cusps in the density of the ordinary fermionic system  are smeared out at any nonzero temperature, the first order phase transition 
in the holographic result survives for small temperatures (the larger the chemical potential, the larger the temperature below which the 
discontinuity persists, see fig.\ \ref{figLLL1}). 
Eventually, for sufficiently large temperatures, in both cases the density becomes monotonically 
increasing with increasing magnetic field, i.e., the transition in the holographic result disappears. 

\end{itemize}

\section{Chiral phase transition}
\label{sec:transition}

Since the chiral phase transition has to be determined numerically in general, the next three subsections are devoted to some limit cases where
the calculation is more transparent. These subsections also serve to discuss the $f\simeq 1$ approximation in the 
chirally broken phase and the possible split of chiral and deconfinement phase transitions.

\subsection{Zero magnetic field}
\label{sec:B0}

In the chirally broken phase at vanishing magnetic field $b=0$, the location of the tip of the connected flavor branes is 
given by eq.\ (\ref{u0smallb}),
\be \label{u00}
u_0(b=0) = \frac{16\pi}{\ell^2}\left[\frac{\Gamma\left(\frac{9}{16}\right)}{\Gamma\left(\frac{1}{16}\right)}\right]^2 \simeq 
\frac{0.52}{\ell^2} \, .
\ee
We recall that here we have employed the approximation $f\simeq 1$. In this case, we see that there is a unique solution for $u_0$ for any 
given $\ell$. This solution can become arbitrarily small. We need to ensure, however, that $u_0>u_T$ in order to 
avoid the artifact of the flavor branes hanging farther down than they are 
allowed to by the geometry, see fig.\ \ref{figcylinders}. With the result (\ref{u00}) and the definition of $u_T$ in eq.\ (\ref{defuT})
this condition is equivalent to 
$t <\frac{3}{\ell\,\pi^{1/2}}\Gamma\left(\frac{9}{16}\right)/\Gamma\left(\frac{1}{16}\right)\simeq 0.173/\ell$, which yields a temperature
limit at $b=0$ for the applicability of our approximation. 

In the full treatment, there is a critical value for the separation $\ell$ 
above which there is no solution for $u_0$ (the branes must be disconnected then). For separations smaller than this maximal value 
there are in fact two solutions, one of which is unstable \cite{Bergman:2007wp} and which approaches $u_T$ for $\ell\to 0$ (i.e., when the connected 
flavor branes are very close together they stretch down almost to the horizon). This unstable solution does not exist in the $f\simeq 1$ 
approximation, where the unique solution is an approximation to the stable solution of the full calculation.

At zero magnetic field we have $\jmath=\eta=0$ and thus the free energy of the chirally broken phase (\ref{OmegaB}) becomes  
\be\label{OmUb0}
\Omega_\cup(b=0) = {\cal N}\int_{u_0}^\infty du\, u^{5/2}\frac{u^4}{\sqrt{u^8-u_0^8}} \, ,
\ee
with $u_0$ given by eq.\ (\ref{u00}).

In the chirally symmetric phase, the equations of motion in the $b=0$ limit are obtained by setting $b=a_3=0$ in eqs.\ (\ref{EOMXs}).
This yields a simple differential equation for $a_0$, which, when evaluated at $u=\infty$, 
relates the integration constant $C$ to the chemical potential,
\be \label{muC}
\mu = \int_{u_T}^\infty du\, \frac{C}{\sqrt{u^5+C^2}} \, .
\ee
The free energy can be obtained from eq.\ (\ref{DBICS}). Using the equation of motion for $a_0$ we have 
\be \label{OmIIb0}
\Omega_{||}(b=0) = {\cal N}\int_{u_T}^\infty du\, \frac{u^5}{\sqrt{u^5+C^2}} \, .
\ee
The chiral phase transition is now obtained by finding the zero of the free energy difference
\be
\Delta\Omega \equiv \Omega_{||}-\Omega_\cup \, .
\ee
(While each of the free energies is divergent, their difference is finite.) 
Even in the case $b=0$, the zero of $\Delta \Omega$ has to be found numerically in general.
Our result for zero (and nonzero) magnetic fields is shown in the next subsection in the lower panel of fig.\ \ref{figcuts}.
For vanishing chemical potential we find the analytic result 
\be
\frac{\Delta\Omega(b=\mu=0)}{\cal N} = \frac{2}{7}\left[u_0^{7/2}\pi^{1/2}
\frac{\Gamma\left(\frac{9}{16}\right)}{\Gamma\left(\frac{1}{16}\right)}-u_T^{7/2}\right] \, ,
\ee
which, using eq.\ (\ref{u00}), yields the critical temperature 
\be \label{tcell}
t_c(b=\mu=0) = \frac{3}{\ell\,\pi^{3/7}}
\left[\frac{\Gamma\left(\frac{9}{16}\right)}{\Gamma\left(\frac{1}{16}\right)}\right]^{8/7}\simeq \frac{0.14}{\ell} \, .
\ee
This critical temperature is close to, but still below the upper limit for our approximation discussed above. Our approximate value 
deviates from the full result by about 10\% (see fig.\ 6 in ref.\ \cite{Horigome:2006xu}). We can use our result to estimate for which 
separations $L$ there is a deconfined, chirally broken phase. This phase occurs if $T_c=t_c \ell/L$ is larger than the critical temperature
for deconfinement $T_{c,{\rm deconf.}} = M_{\rm KK}/(2\pi)$. Consequently, the critical $L$ below which a deconfined chirally 
broken phase exists, is $L_c \simeq 0.27 \pi/M_{\rm KK}$ (compared to $L_c\simeq 0.31\pi/M_{\rm KK}$ in the full calculation \cite{Aharony:2006da}).

\subsection{Zero temperature}
\label{sec:T0}

At zero temperature, we can compute the critical chemical potential for vanishing $b$ as well as for asymptotically large $b$ analytically.
Since in our $f\simeq 1$ approximation the chirally broken phase does not depend on temperature, the location of the tip of the connected 
branes $u_0$ and the free energy at $b=t=0$ are simply given by the results of the previous subsection, eqs.\ (\ref{u00}) and (\ref{OmUb0}). 
In the chirally symmetric phase, the value of the constant $C$ at $t=0$ can be determined from eq.\ (\ref{muC}),
\be
C^{2/5} = \frac{\mu\sqrt{\pi}}{\Gamma\left(\frac{3}{10}\right)\Gamma\left(\frac{6}{5}\right)} \, .
\ee
The corresponding free energy is given by inserting this value and $u_T=0$ into eq.\ (\ref{OmIIb0}). As a result, the difference in free energies 
becomes
\be
\frac{\Delta\Omega(b=t=0)}{\cal N} = \frac{2}{7}\left[u_0^{7/2}\pi^{1/2}\frac{\Gamma\left(\frac{9}{16}\right)}{\Gamma\left(\frac{1}{16}\right)}
-C^{7/5}\frac{\Gamma\left(\frac{3}{10}\right)\Gamma\left(\frac{6}{5}\right)}{\pi^{1/2}}\right] \, , 
\ee
which yields the critical chemical potential 
\be \label{muc0}
\mu_c(b=t=0) = \frac{16\pi^{11/14}
\left[\Gamma\left(\frac{3}{10}\right)\Gamma\left(\frac{6}{5}\right)\right]^{5/7}}{\ell^2}
\left[\frac{\Gamma\left(\frac{9}{16}\right)}{\Gamma\left(\frac{1}{16}\right)}\right]^{16/7}
 \simeq \frac{0.44}{\ell^2}\, .
\ee  
Since at $t=0$  we have $f=1$, this result is exactly the same as in fig.\ 6 of ref.\ \cite{Horigome:2006xu}.

At asymptotically large magnetic field,  $\sinh y_\infty$ diverges and thus eqs.\ (\ref{u0eta2}) implies $\eta=0$ while from eq.\ (\ref{u0eta1})
we obtain   
\be \label{u0B}
u_0(b\to\infty) = \frac{16\pi}{\ell^2}\left[\frac{\Gamma\left(\frac{3}{5}\right)}{\Gamma\left(\frac{1}{10}\right)}\right]^2 \simeq 
\frac{1.23}{\ell^2} \, .
\ee
For the free energy in the chirally broken phase we insert $y_\infty \to \infty$ into eq.\ (\ref{OmegaB}), while in the chirally 
symmetric phase we use $z_\infty\to \infty$ in eq.\ (\ref{OmegaXs}). Consequently,
\be \label{dOmlargeb}
\frac{\Delta\Omega(t=0,b\to\infty)}{\cal N} = \frac{b}{2}\left[u_0^2\pi^{1/2}\frac{\Gamma\left(\frac{3}{5}\right)}{\Gamma\left(\frac{1}{10}\right)}
-\frac{9}{4}\mu^2\right] \, , 
\ee
which yields the critical chemical potential at asymptotically large $b$,
\be \label{mucasymp}
\mu_c(t=0,b\to\infty) = \frac{32\pi^{5/4}}{3\ell^2}\left[\frac{\Gamma\left(\frac{3}{5}\right)}{\Gamma\left(\frac{1}{10}\right)}\right]^{5/2}
\simeq \frac{0.43}{\ell^2} \, .
\ee
Hence the critical chemical potentials at $b=0$ and $b\to\infty$ are almost identical. This is confirmed by the numerical 
result for arbitrary $b$ which is presented in sec.\ \ref{sec:IMC} in fig.\ \ref{figzoomin}. 

It is important to 
specify that we compare the free energies of the two phases at a fixed value of the microscopic magnetic field $B$, not the externally 
applied field $H$. Had we fixed $H$, we would have had to perform a Legendre transformation of our free energy, as done in 
ref.\ \cite{Rebhan:2008ur}. In general, the physical context dictates which field must be held fixed. Here we are mostly interested in a comparison 
with previous NJL calculations, where $B$ is fixed.

\subsection{Zero chemical potential: when do chiral and deconfinement transitions split?}
\label{sec:mu0}

In the chirally broken phase at $\mu=0$, eq.\ (\ref{jmin}) implies $\jmath=0$ and thus eqs.\ (\ref{u0eta}) yield $\eta=0$ and 
\be
\frac{\ell}{2} = \int_{u_0}^\infty du\,\frac{\sqrt{u_0^8+b^2u_0^5}}{u^{3/2}\sqrt{u^8+b^2u^5-(u_0^8+b^2u_0^5)}} \, ,
\ee
which is an equation for $u_0$, to be solved numerically. One finds that $u_0$ increases monotonically 
with $b$ and saturates at a finite value for asymptotically large $b$. This value can be computed analytically and is given by
eq.\ (\ref{u0B}). (Once we let $b\to\infty$, the values of $\eta$ and $u_0$ at the minimum of the free energy become independent of $\mu$.)

In eq.\ (\ref{u00}) we have seen that the tip of the branes can be lifted
by decreasing their asymptotic separation $\ell$. Now we see that a magnetic field has a similar effect: for a fixed separation $\ell$ 
a magnetic field increases $u_0$ from the value (\ref{u00}) at $b=0$ to the value (\ref{u0B}) for $b=\infty$. 
Large values of $u_0$ tend to favor the chirally broken phase: see for instance
eq.\ (\ref{tcell}) which shows that decreasing the asymptotic separation $\ell$ (and thus increasing $u_0$) 
increases the critical temperature $t_c$. Therefore, 
a magnetic field seems to favor chiral symmetry breaking, which is in accordance with the expectation of MC. We will discuss
this in more detail in the next subsection, where we show that one cannot naively transfer this expectation to the 
case of nonzero chemical potential. 

The free energy of the chirally broken phase (\ref{OmegaB}) becomes with $\jmath=\eta=0$
\be
\Omega_\cup(\mu=0) = {\cal N}\int_{u_0}^\infty du\,u^{3/2}\frac{u^5+b^2u^2}{\sqrt{u^8+b^2u^5-(u_0^8+b^2u_0^5)}} \, , 
\ee
while for the symmetric phase we have
\be
\Omega_{||}(\mu=0) = {\cal N}\int_{u_T}^\infty du\,\sqrt{u^5+b^2u^2} \, .
\ee
The phase transition for arbitrary $b$ must be determined numerically. 
\FIGURE[t]{\label{figsplit}
{ \centerline{\def\epsfsize#1#2{0.8#1}
\epsfbox{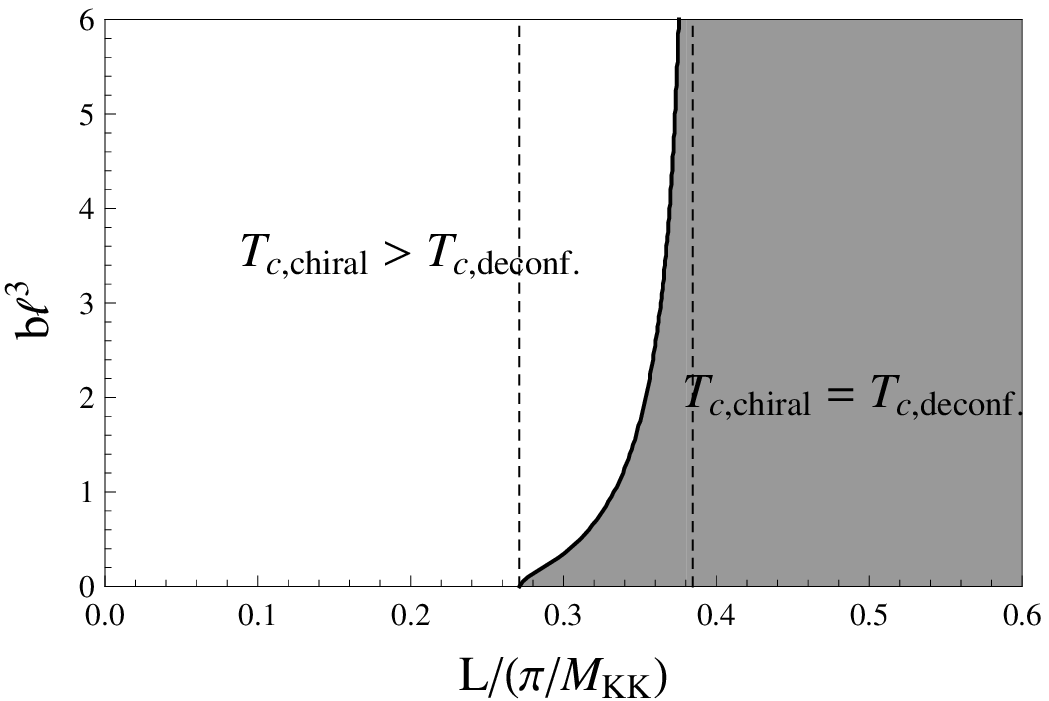}
}
}
\caption{Region in the $b$-$L$ parameter space where the chiral and deconfinement phase transitions split (white area) and where 
they coincide (gray area). For intermediate separations $L$ of the flavor branes -- between the two dashed lines -- a sufficiently 
large magnetic field $b$ induces a split in the phase transitions. (While in all other plots it is understood implicitly that $b$ is rescaled
by $\ell=L/R$, we have written this rescaling explicitly here since $L$ also appears on the horizontal axis.)  
 }  
}
The resulting critical line for zero (and nonzero) chemical potential is presented
in the next subsection in the middle panel of fig.\ \ref{figcuts}. Using the complete function $f$, this line has been computed in  
refs.\ \cite{Bergman:2008sg,Johnson:2008vna}. Here we continue with an analytic result for asymptotically large $b$ [for $b=0$
the result is given in eq.\ (\ref{tcell})]. In this case,
the difference of free energies becomes
\be
\frac{\Delta\Omega(\mu=0,b\to\infty)}{\cal N} = \frac{b}{2}\left[u_0^2\pi^{1/2}
\frac{\Gamma\left(\frac{3}{5}\right)}{\Gamma\left(\frac{1}{10}\right)}-u_T^2\right] \, ,
\ee
and thus, using the asymptotic expression for $u_0$ (\ref{u0B}), the critical temperature is
\be \label{tcasymp}
t_c(\mu=0,b\to\infty) = \frac{3}{\ell \pi^{3/8}}\left[\frac{\Gamma\left(\frac{3}{5}\right)}{\Gamma\left(\frac{1}{10}\right)}\right]^{5/4}
\simeq \frac{0.19}{\ell} \, .
\ee
Again, we can now determine the critical value of the separation of the flavor branes below which the chiral and deconfinement phase transitions 
split. We find $L_c\simeq 0.38\pi/M_{\rm KK}$.
This critical value is larger than without magnetic field. We can determine $L_c$ for arbitrary $b$ numerically to obtain a phase diagram with 
regions where the phase transitions coincide and where they don't, see fig.\ \ref{figsplit}. 

The geometric picture is as follows. Suppose we set $\mu=0$ and
choose the temperature slightly (infinitesimally) larger than the deconfinement phase transition (this transition is, in the probe brane 
approximation $N_c\gg N_f$, completely determined by $M_{\rm KK}$ and independent of all quantities 
on the flavor branes such as $\mu$, $b$, and the separation $\ell$). It is now possible to choose the asymptotic separation of the flavor branes 
such that they can connect (meaning that the connected branes constitute the ground state of the system). To this end, the separation has to be 
sufficiently small. There is a regime of asymptotic separations (to the left of the left dashed line in fig.\ \ref{figsplit})
where the flavor branes always connect, even for zero magnetic fields; there is another regime 
(to the right of the right dashed line) where the flavor branes never connect, even for asymptotically large magnetic
fields; and there is a regime (between the two dashed lines) where the larger the magnetic field the farther apart we can put the connected
flavor branes.

Another way to read this figure is to consider the horizontal axis as a parameter that interpolates between different 
dual field theories. Large separations $L$ correspond to large-$N_c$ QCD, where the gluon dynamics becomes important. In this case, the magnetic
field cannot induce a split of chiral and deconfinement phase transitions. Small separations $L$ correspond to an NJL-like model, where
the chiral and gluon dynamics decouple. While sufficiently small separations split the phase transitions for arbitrary magnetic field, there
is an interesting intermediate regime where only a sufficiently large magnetic field induces a split.

Although fig.\ \ref{figsplit} has been obtained for vanishing chemical potential, the conclusions are easy to generalize to all values of $\mu$.
As we shall see in the next subsection, the critical temperature for the chiral phase transition is maximal for $\mu=0$. Consequently, 
if and only if there is a split in the phase transitions at $\mu=0$ there is also a split for a finite regime of nonzero $\mu$.

\subsection{General results: inverse magnetic catalysis and comparison to NJL}
\label{sec:IMC}

We can now compute the free energy difference between chirally broken and chirally symmetric phases for all $b$, $t$, and $\mu$.
The resulting chiral phase transition is presented in figs.\ \ref{figzoomin} -- \ref{fig3D}, where figs.\ \ref{figzoomin} and \ref{figcuts} 
are two-dimensional cuts through the three-dimensional phase diagram shown in fig.\ \ref{fig3D}. In fig.\ \ref{figzoomin} and in the 
three-dimensional plot we show, in addition to the chiral phase transition, the ``Landau level'' transition discussed in sec.\ \ref{sec:Tzero2}. 
(In fig.\ \ref{figcuts} we have omitted this transition in order to keep the plots simple.)

At the chiral phase transition line, the baryon number density increases
from a purely topological contribution, which can be viewed as a stack of $\pi^0$ domain walls \cite{Thompson:2008qw,Son:2007ny}, to one that is carried
by chirally symmetric quarks.
For simplicity, our calculation does not take into account ``normal'' baryonic matter in the chirally broken phase, which in the Sakai-Sugimoto model can be
represented by D4 branes wrapped on the $S_4$ within the D8 branes.
To get an idea about the possible onset of a normal baryon density we have computed the zero-temperature constituent quark mass which, in the setup 
with D4 branes, is $m(\mu,b) = u_0(\mu,b)/3$ \cite{Bergman:2008qv,Bergman:2007wp,Rozali:2007rx} (in the same dimensionless units as $\mu$), 
and have plotted the (thin dotted) 
line $\mu=u_0(\mu,b)/3$ in fig.\ \ref{figzoomin}. 
For $\mu>m(\mu,b)$, an admixture of normal baryonic matter may occur
in the broken phase. 
To obtain the actual transition to the phase where 
topological and normal baryonic matter coexist, and where it
again ends, a consistent calculation including the effect of the baryon mass on the embedding of the flavor 
branes would have to be performed, which is however beyond the scope of
this paper.

The most interesting observations resulting from our present calculations
are as follows.

\FIGURE[t]{\label{figzoomin}
{ \centerline{\def\epsfsize#1#2{0.9#1}
\epsfbox{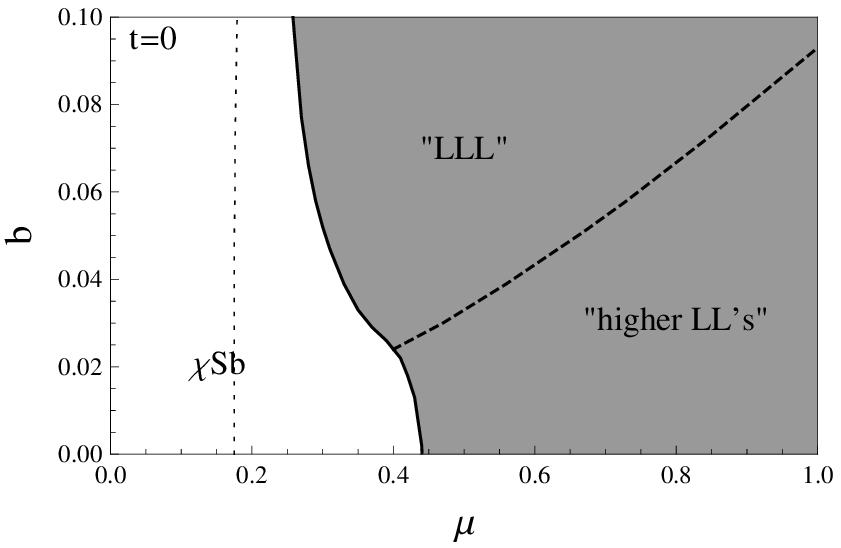}
\def\epsfsize#1#2{0.86#1}
\epsfbox{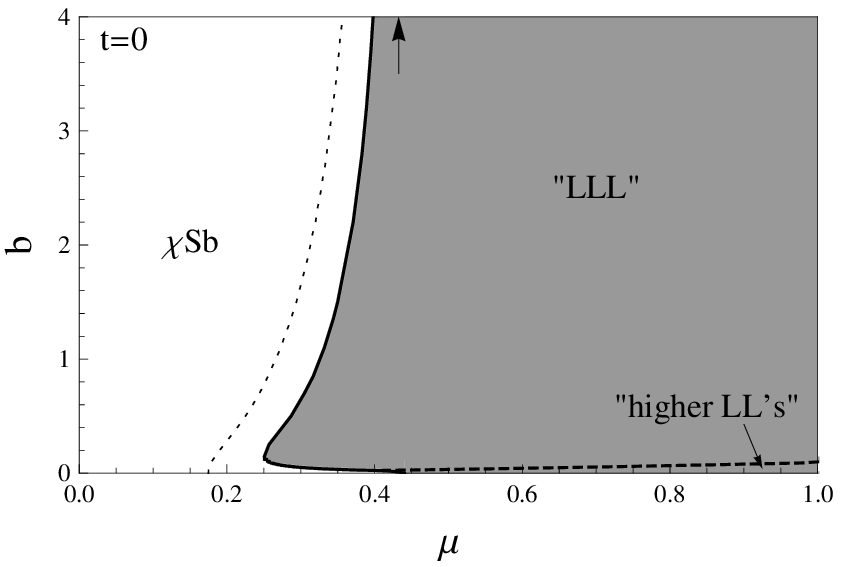}
}
\vspace*{-2mm}
}
\caption{Chiral phase transition (solid line) at zero temperature in the $b$-$\mu$ plane on a small (left) and a large (right)
scale of the magnetic field. The first-order critical line divides the chirally broken phase ($\chi$Sb, white area) from the 
chirally restored phase (gray area). It starts at the value (\ref{muc0}) for $b=0$ and approaches the value (\ref{mucasymp})
for asymptotically large $b$ (marked by an arrow in the right panel). The behavior in between is one of our main results since it shows that
for finite chemical potential the presence of a small magnetic field disfavors chiral symmetry breaking. 
The dashed line marks the discontinuity of the quark density in the chirally symmetric phase. This transition is reminiscent of 
a Landau level transition, as indicated in the figure, see sec.\ \ref{sec:Tzero2} for a discussion of this line. The thin dotted line 
$\mu=u_0(\mu,b)/3$ marks the potential onset of ``normal'', chirally broken 
baryonic matter to the right of this line (which is not included in our calculation).
 }  
}

\FIGURE[t]{\label{figcuts}
{ 
\centerline{
\def\epsfsize#1#2{0.92#1}
\epsfbox{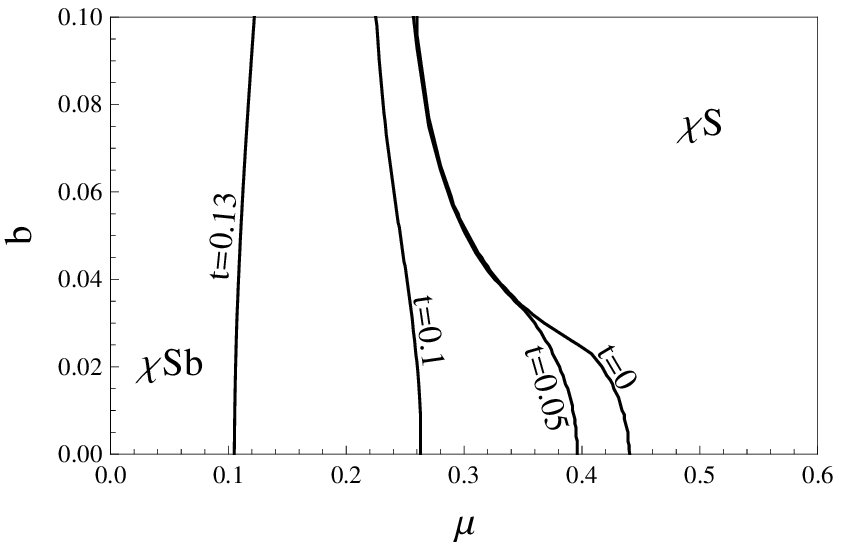}}\\[1.5ex]
\centerline{
\def\epsfsize#1#2{0.74#1}
\epsfbox{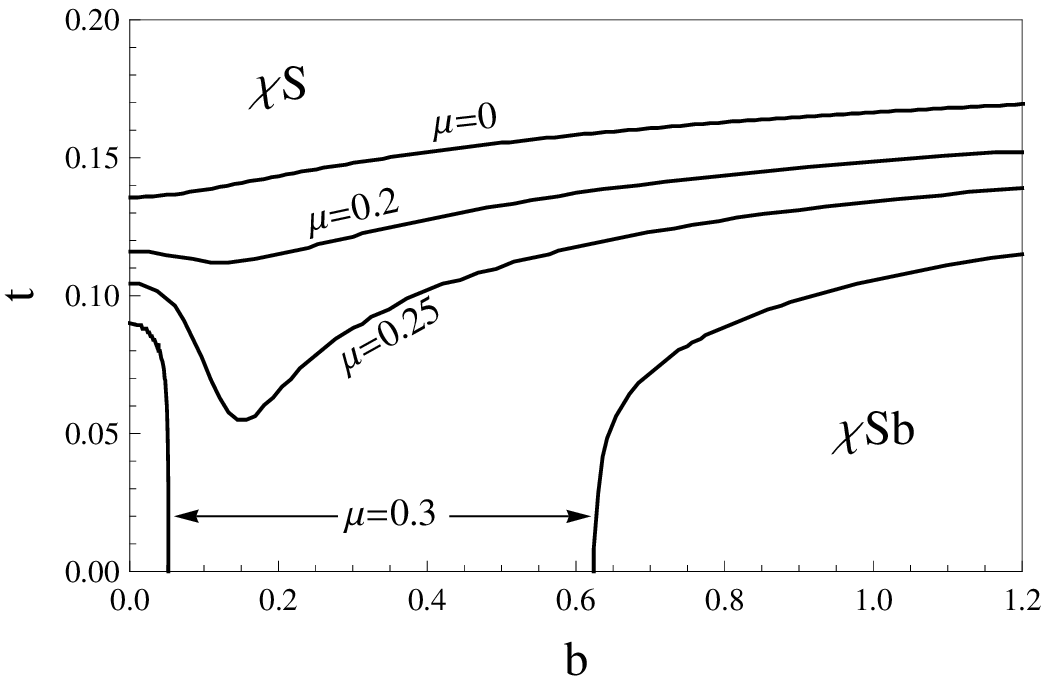}}\\[1.5ex]
\centerline{
\def\epsfsize#1#2{0.93#1}
\epsfbox{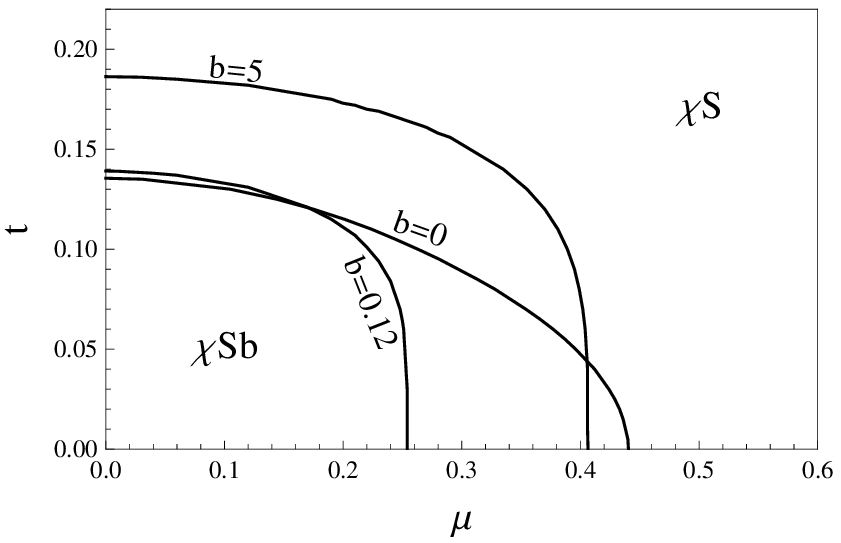}}
}
\vspace*{-5mm}
\caption{Transition between the chirally broken ($\chi$Sb) and chirally symmetric ($\chi$S) phases in the $b$-$\mu$, $t$-$b$, and $t$-$\mu$ 
planes for several fixed temperatures, chemical potentials, and magnetic fields, respectively (i.e., all panels are two-dimensional cuts through the 
three-dimensional phase diagram shown in 
fig.\ \ref{fig3D}). All lines are first-order phase transition lines. The quantities $b$, $\mu$, $t$ are dimensionless and related to the 
dimensionful counterparts by appropriate factors of ($2\pi$ times) the string tension $\alpha'$ and the curvature radius $R$; 
moreover, we have set the asymptotic separation $\ell$ of the flavor branes to 1, the $\ell$ dependence is recovered by 
replacing $b\to b\ell^3$, $\mu\to\mu\ell^2$, $t\to t\ell$. For simplicity we have omitted the ``Landau level'' transition lines shown in 
fig.\ \ref{figzoomin}.
 }  
}

{\it Inverse magnetic catalysis.--}
In fig.\ \ref{figzoomin} we see that by increasing the magnetic field up to $b\lesssim 0.2/\ell^3$ 
at zero temperature and finite chemical potential, the chirally broken phase becomes {\it less} favorable. As discussed in the 
introduction, one might have expected a magnetic field to favor chiral symmetry breaking due to ``magnetic catalysis'' (MC). 
We term the observed opposite effect ``inverse magnetic catalysis'' (IMC). 
For larger magnetic fields, $b\gtrsim 0.2/\ell^3$, the phase transition line bends back and the magnetic field tends to favor the chirally 
broken phase, as expected from MC.\footnote{Had we restricted ourselves to isotropic configurations without a supercurrent, the phase transition line would show a somewhat enhanced IMC and then a weaker MC at large $b$, asymptoting to the smaller value $\mu=\frac{\sqrt{3}}{2}\mu_c(t=0,b\to\infty)$.} 
 Note that the two
opposite effects occur on different scales of the magnetic field: the right panel of fig.\ \ref{figzoomin} shows a large
scale on which the phase transition line approaches its asymptotic value in accordance with MC; on this scale the opposite IMC 
at small magnetic fields is barely visible. 

The IMC becomes less pronounced for nonzero temperatures but exists up to $t\lesssim 0.1$, as we see in the upper panel of fig.\ \ref{figcuts}. 
It manifests itself also in the middle and lower panels of this figure. For instance, in the middle panel we see a monotonically increasing critical 
temperature for $\mu=0$. For nonzero $\mu$, however, the critical temperature becomes non-monotonic. 
There is even an intermediate range of $\mu$, here shown for $\mu=0.3$, for which sufficiently cold matter is chirally broken at small and 
at large magnetic field, but not in between.

\FIGURE[t]{\label{fig3D}
{ \centerline{\def\epsfsize#1#2{0.4#1}
\epsfbox{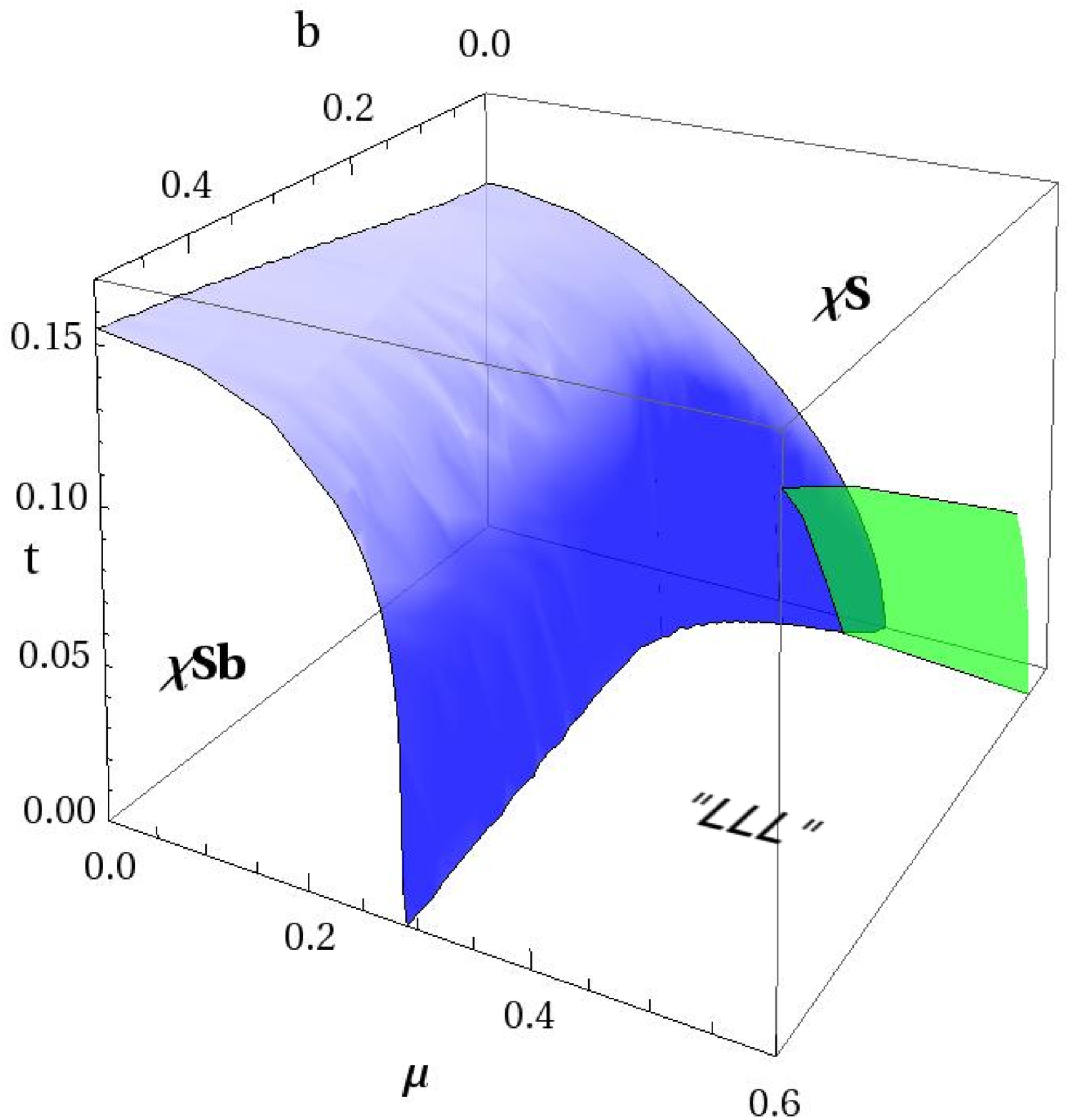}
}
}
\caption{Chiral phase transition (large, blue surface) in the parameter space of magnetic field $b$, temperature $t$ and chemical
potential $\mu$. Two-dimensional cuts through this space (sometimes with larger $b$ than shown here) are presented in fig.\ \ref{figcuts}.
We have indicated by ``LLL'' that, at $t=0$, the quark number density in the chirally symmetric ($\chi$S) phase on the large-$b$ side of the 
(small, green) critical surface is identical to that of free fermions in the lowest Landau level. In general, this critical surface indicates
a discontinuity in the quark number density. 
 }  
}

{\it Comparison with NJL calculations.--} It is interesting to compare our results with corresponding NJL 
calculations \cite{Ebert:1999ht,Inagaki:2003yi,Boomsma:2009yk,Fayazbakhsh:2010bh}. For instance, our fig.\ \ref{figcuts} and 
fig.\ 4 in ref.\ \cite{Inagaki:2003yi} show an amazing agreement in the chiral phase transition lines throughout the $t$-$b$-$\mu$ space;
in particular, both results show IMC for moderate magnetic 
fields.\footnote{Ref.\ \cite{Yamada:2009qb} claims that the Sakai-Sugimoto model does {\it not} 
agree with NJL results; but in this reference the CS contribution was ignored.} In QCD, it is expected that 
part of the normal chirally symmetric phase is replaced by a color superconductor \cite{Alford:2007xm}.  
From the results of the NJL calculation in 
Ref.\ \cite{Fayazbakhsh:2010bh} one can read off that also in this case IMC is present for small temperatures. 

For $t=0$, there are interesting similarities between our fig.\ \ref{figzoomin} with 
fig.\ 4 in ref.\ \cite{Ebert:1999ht} as well as with fig.\ 2 in ref.\ \cite{Boomsma:2009yk}. 
Namely, in the left panel of fig.\ \ref{figzoomin} we see that roughly at 
the point where the (dashed) critical line ends at the (solid) phase transition line, the latter strongly bends to the left. Such a structure is 
also seen in the NJL results, where the critical line marks the onset of the first Landau level. In the NJL model, more critical lines end at the 
chiral phase transition line, in principle one for each additionally occupied Landau level, giving rise to de Haas-van Alphen oscillations in the transition line. These additional lines and corresponding oscillations 
are absent in our approach, suggesting a separated ``LLL'' from a continuum of ``higher LL's'', as indicated in fig.\ \ref{figzoomin}. 

Besides the absence of higher Landau levels, there are more differences to the NJL model. In particular, all our phase transitions are of 
first order. For the chiral phase transition this is obvious from the geometric point of view since there is a discontinuous transition 
from connected to disconnected flavor branes. More physically speaking, there is a jump in the density across all phase transition lines 
shown in the plots. The NJL model, however, shows  first-order and second-order phase transition lines in the chiral limit \cite{Inagaki:2003yi}.

{\it Discussion.--} Why does dense (holographic) matter not behave as suggested by MC and rather shows a complicated mixture of MC and IMC? Naively, 
one might think that there are two ``forces'' acting on the chiral condensate: the magnetic field tends to effectively increase the 
particle-antiparticle coupling, as explained in the original works about 
MC \cite{Gusynin:1994re,Gusynin:1994va,Gusynin:1994xp,Gusynin:1995gt},
and thus work in favor of chiral symmetry breaking; the chemical potential tends to split particles from antiparticles which puts a stress
on a particle-antiparticle pair, thus working against chiral symmetry breaking. And indeed, our results show instances where these two effects, 
considered separately, can be observed. Without $\mu$, the critical temperature is increased by $B$, 
see middle panel of fig.\ \ref{figcuts}; without $B$, the critical temperature is decreased by $\mu$, see lower 
panel of fig.\ \ref{figcuts}. This suggests that if we increased $B$ at fixed $\mu$, only one of the ``forces'' would be at work, 
favoring chiral symmetry breaking. However, in fig.\ \ref{figzoomin} we have seen that this is not true in general: 
for a certain range of chemical potentials chiral symmetry is restored upon increasing $B$.

To explain this effect, recall the analogy of magnetically-induced chiral symmetry breaking and superconductivity \cite{Gusynin:1994xp} at
weak coupling. In the case of a superconductor, conventional BCS Cooper pairing between (massless) fermion species whose Fermi surfaces are split 
by a mismatch 
$\delta\mu$ is possible if the pairing gap $\Delta$ is sufficiently large. One can picture this situation as follows. Start from two different 
filled Fermi spheres whose radii differ by $\delta\mu$. For conventional Cooper pairing to happen at zero temperature, the
Fermi surfaces must coincide. To this end, force both Fermi surfaces to the common, average Fermi surface $\mu$. Creating this
fictitious, intermediate state results in a free energy cost $\propto \mu^2\delta\mu^2$. But now pairing yields a gain in free energy 
$\propto \Delta^2\mu^2$.
Consequently, pairing is possible if $\Delta$ is large compared to $\delta\mu$ and breaks down otherwise. Working out the correct prefactors, one
finds that Cooper pairing breaks down for $\delta\mu > \Delta/\sqrt{2}$, which is called the Clogston-Chandrasekhar
relation \cite{Clogston:1962zz,Chandrasekhar}.

We can transfer this picture to the chiral condensate in a strong magnetic field as follows. First we note that we can restrict ourselves to 
the physics of the LLL, since both NJL and holographic results show the strongest IMC in a regime where the higher Landau levels are empty. 
While for the usual superconductor at $\delta\mu=0$ the fermions that ``want''
to pair sit on the two-dimensional surface $k=k_F$ of the Fermi sphere ($k_F$ being the Fermi momentum), the LLL fermions and antifermions 
that ``want'' to form a chiral condensate both sit, at $\mu=0$, on the two-dimensional plane $k_3=0$ perpendicular to the magnetic field 
in the 3-direction.  Now we switch on $\mu$, which in our context is the analogue of $\delta\mu$ because it separates the fermion surface from the 
antifermion surface. As above, we imagine to force the two planes back to their $\mu=0$ position. The resulting energy cost is 
$\propto B\mu^2$. This can be seen from the zero-temperature limit of eq.\ (\ref{Omegafree}) which shows that the LLL contribution is 
$\Omega = -N_cB\mu^2/(4\pi^2)$ which becomes $\Omega=0$ for $\mu=0$. Independent of the precise form of the energy gain due to 
the formation of a chiral condensate -- which is also expected to increase with $B$ -- our first important observation is that the {\it cost} 
increases with $B$. (This is almost as if, in the case of the 
superconductor, both $\delta\mu$ and $\Delta$ increase upon increasing a single parameter.) Therefore, the competition between
the effects of $\mu$ and $B$ is more complicated than naively expected, and we understand why IMC {\it can} happen. Whether it {\it does}
happen depends on the coupling strength as we now explain. 

In the weak-coupling limit, the free energy difference obtained in an NJL model is 
\cite{Gorbar:2009bm,Gorbar:2011ya}, 
\be \label{clog}
\Delta\Omega \propto B\left(\frac{M^2}{2}-\mu^2\right) \, , 
\ee 
where $M$ is the $B$-dependent constituent quark mass. In this case, there is an exact analogy to the 
Clogston-Chandrasekhar relation, namely $\mu>M/\sqrt{2}$, and increasing $B$ at fixed $\mu$ can only increase, never decrease, $\Delta\Omega$
(since $M$ increases with $B$ at weak coupling due to MC).
This shows that IMC is not possible in the weak-coupling limit. Interestingly, 
the free energy difference in our holographic calculation assumes the same form for asymptotically large $B$ if we identify $M=u_0R/(2\pi\alpha')$ 
\cite{Johnson:2008vna,Aharony:2006da}, as we can see from eq.\ (\ref{dOmlargeb}). Therefore, we might speculate that the limit 
of asymptotically large
magnetic fields, where we do not observe IMC, is in some sense equivalent to the weak-coupling limit. 
The reason might be that the magnitude of the constituent quark mass can be
interpreted as a measure for the coupling strength, and in the given limit the constituent mass (squared) is much smaller than $B$. However, as 
eq.\ (\ref{dOmlargeb}) shows, the relation between the condensation energy and the constituent
quark mass involves a more complicated numerical factor compared to the NJL model. (For a comparison with eq.\ (\ref{clog}) we need to 
consider an isotropic condensate, i.e., switch off the supercurrent; even after this modification the prefactor is different.)

Our observation of IMC at smaller magnetic fields suggests that the free energy difference must change qualitatively. The simplest way
to see this is to use the small-$B$ approximation for the chirally broken phase and the ``LLL'' result for the symmetric phase. From 
eqs.\ (\ref{OmegaXs}) and (\ref{OmUb0}) we find with $u_0 \propto M$ 
\be \label{MmuB}
\Delta\Omega \simeq  {\rm const}\times M^{7/2}- \mu^2 B \, ,
\ee
i.e., the condensation energy has dramatically changed while the cost of forming a condensate has remained the same. For small magnetic
fields we have $M={\rm const} + {\cal O}(B^2)$ (see appendix \ref{app:d} for the precise form of this expansion), and thus at some 
value of $B$ the cost exceeds the gain, resulting in IMC. Two comments about eq.\ (\ref{MmuB}) are in order. Firstly, one might question 
the validity of this free energy because we have used an expansion in small $B$, although the ``LLL'' phase exists only at sufficiently large $B$. 
Indeed, to obtain a good approximation to the full result, at least ${\cal O}(B^2)$ terms have to be included, see appendix \ref{app:d}.
However, this does not change the conclusion regarding IMC whose qualitative form is well captured by the above linear approximation.  
Secondly, eq.\ (\ref{MmuB})
shows that our conclusions for a truly four-dimensional field theory have to be taken with some care: $M^{7/2}$ does not have mass dimensions of a 
free energy density; the constant contains the dimensionful factor $M_{\rm KK}^{1/2}$, reflecting the extra dimension in our model. Nevertheless, 
the qualitative agreement with NJL calculations suggests that our observation is of general nature
and may thus also be relevant for QCD.

What is the range of magnetic fields in physical units for which IMC occurs? For a rough estimate let us match the $b=0$ values 
of our critical temperature at $\mu=0$ and our critical chemical potential at $t=0$ to the approximate values from QCD, $T_c\simeq 150\, {\rm MeV}$
and $\mu_{q,c}\sim 400\, {\rm MeV}$ (the former is in fact
a cross-over rather than a critical temperature, as
known from lattice calculations \cite{Aoki:2006br,Aoki:2006we}, while for the latter we only have a comparatively rough idea, see 
e.g.\ refs.~\cite{Rebhan:2003wn,Kurkela:2009gj}).  We can express the dimensionful quantities as
\be
\mu_q = \frac{R^3}{2\pi\alpha'}\frac{\mu\ell^2}{L^2} \, , \qquad T = \frac{t\ell}{L} \, , \qquad |qB| = \frac{R^3}{2\pi\alpha'}\frac{b\ell^3}{L^3} \, ,
\ee
where we have reinstated the electric charge $q$.
This shows that, expressed in terms of the dimensionful model parameters $R^3/(2\pi\alpha')$ and $L$, the different scalings with respect to $\ell$ 
used in our plots arise naturally.\footnote{With $\frac{R^3}{2\pi\alpha'}=\frac{54\pi^2\kappa}{N_c M_{\rm KK}}$ we recover the 
parameters $\kappa$ and $M_{\rm KK}$, whose values are matched to the physical 
pion decay constant and rho meson mass in ref.\ \cite{Sakai:2005yt}; note, however, that we cannot simply use these numerical values since they are 
only meaningful for a maximal separation $L=\frac{\pi}{M_{\rm KK}}$.} The scale for the magnetic field is now found with the help of these
relations,
\be
\frac{|qB|}{b\ell^3} \simeq 5.1\times 10^{19}\,{\rm G}\,\left(\frac{\mu_{q,c}}{400\,{\rm MeV}}\right)\left(\frac{T_c}{150\,{\rm MeV}}\right) \, ,
\ee
where we have inserted our results for $ t_c\ell$ and $\mu_c\ell^2$ from eqs.\ (\ref{tcell}) and (\ref{muc0}), respectively. 
Now we can read off from fig.\ \ref{figzoomin} that, at 
zero temperature, IMC occurs for magnetic fields up to $|qB| \lesssim 1.0\times 10^{19}\,{\rm G}$ and leads to a reduction of the critical 
chemical potential from (the matched value) $\sim 400\,{\rm  MeV}$ down to $\sim 230\,{\rm MeV}$.
The phase that we have identified with the transition into the lowest Landau
level occurs for $|qB| \gtrsim 1.0\times  10^{18} \,{\rm G}$.

Finally, let us elaborate on the comparison to the NJL model calculations. We have seen interesting similarities in the results. However, in 
the NJL works  mentioned so far \cite{Ebert:1999ht,Inagaki:2003yi,Boomsma:2009yk} only isotropic chiral condensates have been considered. 
As a consequence, the chirally broken
phase has, at least at small chemical potential, vanishing baryon number. Only if the quark chemical potential becomes
larger than the constituent quark mass, the baryon number may be nonzero in the broken phase \cite{Ebert:1999ht,Boomsma:2009yk}. This is different
in our holographic calculation. Here we have an anisotropic chiral condensate throughout the chirally broken phase, which manifests itself in the 
nonzero supercurrent $\jmath$ and a nonzero topological baryon number. As discussed above, we have not included ``normal'' homogeneous baryonic 
matter into our calculation. In view of these different kinds of baryonic densities it appears quite remarkable that the holographic and NJL phase 
diagrams look similar. 

In one recent NJL calculation \cite{Frolov:2010wn}, however, a more general ansatz has been considered. And indeed, as in 
our calculation, an anisotropic chiral condensate is found to be favored throughout the chirally broken phase.
Curiously, the resulting phase diagram in fig.\ 1 of ref.\ \cite{Frolov:2010wn} looks less
similar to our result, compared to the phase diagrams in refs.\ \cite{Ebert:1999ht,Inagaki:2003yi,Boomsma:2009yk}, which are obtained 
with an isotropic condensate. We can only speculate whether this might be due to 
the specific choice of the coupling constant. 
For a more reliable comparison to the NJL phase diagram it is crucial to extend our holographic results by including ``normal'' baryonic matter. 

A further possible complication is the so-called ``chiral shift'' in the symmetric phase \cite{Gorbar:2009bm,Gorbar:2010kc}. NJL model
calculations predict this difference in the dispersions of left- and right-handed fermions in the presence of a magnetic field, and it remains
to be seen how it influences the chiral phase transition. In our holographic calculation the chiral shift would correspond to a nonvanishing 
boundary value of $a_3$, which is absent in the symmetric phase.

\section{Conclusions}
\label{sec:conclusions}

We have discussed the effect of a magnetic field on the chiral phase transition in the deconfined phase of the Sakai-Sugimoto model. 
In the probe brane approximation $N_c\gg N_f$ applied here, this chiral transition exists only under certain conditions. 
It does not exist if the asymptotic separation of the flavor D8- and $\overline{\rm D8}$-branes in the compactified extra dimension
is sufficiently large because then chiral and deconfinement phase transitions coincide, independent of the magnitude of the magnetic field. 
We have identified an intermediate region for the separation where a magnetic field induces a split of 
the two transitions, allowing for a chirally broken, deconfined phase. For even smaller separations, chiral and deconfinement 
phase transition are distinct even for vanishing magnetic field (at zero chemical potential). This is the regime of separations 
we have considered in the main part of the paper. It corresponds to an NJL-like model on the field theory side since the chiral dynamics 
completely decouples from the gluon dynamics, and confinement 
becomes irrelevant. In this sense, our results are not of direct relevance to QCD, at least not for the interplay of chiral symmetry breaking
and confinement. They may still be relevant for qualitative features of the chiral phase transition in QCD, in particular at large chemical
potential and low temperature, where gluonic degrees of freedom
are less dominant at finite $N_c$. Moreover our results may 
be of general interest for other (effectively)
relativistic systems with flavor symmetry breaking in a magnetic field, for instance graphene.

We have computed the critical surface of the chiral transition in the three-dimensional parameter space of temperature, chemical 
potential and magnetic field. The most interesting result is observed for small temperatures and a certain intermediate range of 
chemical potentials. In this case, starting from a chirally broken phase at zero magnetic field, a small magnetic field induces symmetry
restoration before chiral symmetry is broken again at large magnetic fields. The tendency of a magnetic field to favor 
chiral symmetry breaking is well known from the so-called {\it magnetic catalysis}. Our observation for dense holographic matter
is more complicated because for small magnetic fields we see the opposite effect, which we have termed {\it inverse magnetic catalysis}.
We have explained this effect in a simple analogy with a superconductor with mismatched Fermi momenta. The essence of this argument is that 
the magnetic field not only enhances the fermion-antifermion coupling and thus the energy {\it gain} from forming a chiral condensate 
but also enhances the free energy {\it cost} needed to form antifermion-fermion pairs in the presence of a chemical potential.

We have also pointed out parallels and differences of our results to previous NJL model calculations. Most nontrivial features of our results, 
such as the inverse magnetic catalysis in certain regions of the 
phase diagram, can be observed in an NJL model as well, supporting the interpretation of this specific limit
of the Sakai-Sugimoto model as a holographic strong-coupling, non-local version of NJL. 
In accordance with Ref.~\cite{Lifschytz:2009sz},
we have found that the Sakai-Sugimoto model shows indications
of a Landau level structure. This is suggested by the dependence of the quark density on the magnetic field which we have compared in detail
with the corresponding density in a usual particle picture. 
There is a first-order phase transition within the chirally restored phase
whose location with respect to the chiral phase transition is comparable
to that of
the (second-order) transition into the lowest Landau level in the NJL model.
However, in the Sakai-Sugimoto model there are no
further de Haas-van Alphen oscillations and no
additional transitions corresponding to the
higher Landau levels of the particle picture, which  
seem to be replaced by a continuum of states in the holographic model.

Our work opens several directions for future projects. A straightforward extension is to take into account the full effect of the curved
geometry which becomes important for large temperatures (we have considered the $f(u)\simeq 1$ approximation in the chirally broken 
phase which is a considerable technical simplification). Moreover, 
one should include ``conventional'' baryonic matter in the chirally broken phase,
in addition to the 
baryon number induced by an anisotropic chiral condensate. We have discussed that such an extension is interesting in view of a more
detailed comparison to the NJL model. It would also be interesting to get a deeper understanding of the apparent Landau level structure 
in the Sakai-Sugimoto model, in particular it would be important to understand in which sense this structure is a strong-coupling version of the 
usual discrete Landau levels.

\acknowledgments

This work has been supported by the Austrian science foundation FWF under
project no.\ P22114-N16.
We also 
thank D.\ Boer, A.\ Gynther, K.\ Klimenko, G.\ Lifschytz, V.\ Miransky, I.\ Shovkovy, and S.\ Stricker for valuable discussions and comments.

\appendix 

\section{Semi-analytic solution for $f\simeq 1$ in the chirally broken phase}
\label{app:broken}

Here we explain the solution of the equations (\ref{EOM2}). We first take the pairwise ratio of these equations,
\begin{subequations} \label{EOM3}
\bea
\frac{a_0'}{a_3'} &=& \frac{3ba_3+c}{3ba_0+d} \, , \label{EOM31} \\
\frac{a_0'}{u^3x_4'} &=& \frac{3ba_3+c}{k} \, , \label{EOM32} \\
\frac{a_3'}{u^3x_4'} &=& \frac{3ba_0+d}{k} \, . \label{EOM33}
\eea
\end{subequations}
Evaluating eq.\ (\ref{EOM31}) at the point $u=u_0$ yields [using the boundary conditions (\ref{boundarybroken})]
\be
c=0 \, . 
\ee
With the definition of $\eta$ in eq.\ (\ref{eta}), eq.\ (\ref{EOM2x4}) at $u=u_0$ yields
\be \label{b}
k = \frac{u_0^{3/2}\sqrt{u_0^5+b^2u_0^2}}{\sqrt{1+\eta^2}} \, .
\ee
Here we have used that, if $\eta\neq 0$, we must have $a_3'(u_0)=\infty$ due to the boundary condition 
$x_4'(u_0)=\infty$ and the definition (\ref{eta}). Now we rewrite eq.\ (\ref{EOM31}) as $3ba_0a_0'+da_0'=3ba_3a_3'$ and 
integrate this equation to obtain
\be \label{kappa1}
\frac{3}{2}ba_0^2+da_0 = \frac{3}{2}ba_3^2 +\kappa \, ,
\ee
with another integration constant $\kappa$. Evaluating this at $u=\infty$ and using the boundary conditions in eq.\ (\ref{boundarybroken}), 
yields $\kappa$ as a function of $d$,
\be \label{kappa}
\kappa = \frac{3}{2}b(\mu^2-\jmath^2)+d\mu \, .
\ee
We shall return to the determination of the integration constants below. Now we first rewrite the differential equations.
To this end we insert $u^3x_4'$ from eq.\ (\ref{EOM32}) into eqs.\ (\ref{EOM2a0}) and (\ref{EOM2a3}). The resulting
equations are equivalent to 
\begin{subequations} \label{EOM6}
\bea
\partial_y a_0 &=& a_3 \, , \\
\partial_y a_3 &=& a_0 + \frac{d}{3b} \, , 
\eea
\end{subequations}
and a condition for the new variable $y$, 
\be \label{duy}
y' = \frac{3b}{\sqrt{u^5+b^2u^2-6b\kappa-d^2-\frac{k^2}{u^3}}} \, .
\ee
(Remember that the prime always denotes the derivative with respect to $u$; derivatives with respect to $y$ are written explicitly.)
We shall discuss the solution of eqs.\ (\ref{EOM6}) below; first we use the result for $y'$ to obtain the solution for $x_4$ and 
more relations between the integration constants. From eq.\ (\ref{EOM32}) we obtain
\be \label{u3}
u^3 x_4' = \frac{k}{3b}\frac{a_0'}{a_3} = \frac{k}{3b}y' \, .
\ee
Consequently, the condition that $x_4'$ diverge at $u=u_0$ implies that $y'$ also has to diverge at $u=u_0$. Hence, the 
denominator  on the right-hand side of eq.\ (\ref{duy}) has to vanish for $u=u_0$ which implies
\be \label{dkappa}
6b\kappa + d^2 = \frac{\eta^2}{1+\eta^2}(u_0^5+b^2u_0^2) \, , 
\ee
and thus $y(u)$ is given by 
\be \label{yuapp}
y(u) = 3b\sqrt{1+\eta^2}\int_{u_0}^u \frac{v^{3/2}dv}{\sqrt{g(v)}} \, ,
\ee
with $g(v)$ defined in eq.\ (\ref{defg}). From eqs.\ (\ref{u3}) and (\ref{b}) we immediately obtain 
the final solution for $x_4$,
\be \label{x4uapp}
x_4(u) = u_0^{3/2}\sqrt{u_0^5+b^2u_0^2}\int_{u_0}^u\frac{dv}{v^{3/2}\sqrt{g(v)}} \, . 
\ee
Inserting this expression into the boundary condition for the asymptotic separation $\ell$, see eq.\ (\ref{boundx4}), yields 
one of the two equations relating $\eta$ and $u_0$, 
\be \label{u0eta1app}
\frac{\ell}{2} = u_0^{3/2}\sqrt{u_0^5+b^2u_0^2}\int_{u_0}^\infty\frac{du}{u^{3/2}\sqrt{g(u)}} \, .
\ee
The differential equations (\ref{EOM6}) are solved by
\begin{subequations}
\bea
a_0(y) &=& c_1 \cosh y + c_2\sinh y -\frac{d}{3b} \, , \\
a_3(y) &=& c_1\sinh y + c_2 \cosh y \, .
\eea
\end{subequations}
The boundary condition $a_3(u_0)=0$ becomes $a_3(y=0)=0$ which implies
\be
c_2=0 \, .
\ee
Then, with $y_\infty\equiv y(u=\infty)$ we have the boundary conditions $a_0(y_\infty)=\mu$ and $a_3(y_\infty)=\jmath$ and thus
\be \label{coth}
c_1 = \frac{\mu +\frac{d}{3b}}{\cosh y_\infty} = \frac{\jmath}{\sinh y_\infty} \, .
\ee
Consequently, we arrive at the final solution for the gauge fields, 
\begin{subequations} \label{ayapp}
\bea
a_0(y) &=& \mu + \frac{\jmath}{\sinh y_\infty} (\cosh y -\cosh y_\infty)  \, , \\
a_3(y) &=& \frac{\jmath}{\sinh y_\infty}\sinh y  \, .  
\eea
\end{subequations}  
The second remaining equation to determine $u_0$ and $\eta$ is obtained by rewriting eq.\ (\ref{eta}) with the help of eq.\ (\ref{u3}),
\be
\eta = \frac{3bu_0^{3/2}}{k}\partial_y a_3\Big|_{y=0} = \frac{3bu_0^{3/2}}{k}\frac{\jmath}{\sinh y_\infty} 
\, ,
\ee
where in the last step we have used eq.\ (\ref{ay2}). Inserting $k$ from eq.\ (\ref{b}) yields 
\be \label{u0eta2app}
\frac{\jmath}{\sinh y_\infty} = \frac{\sqrt{u_0^5+b^2u_0^2}}{3b}\frac{\eta}{\sqrt{1+\eta^2}} \, .
\ee
As a check, this relation can also be obtained as follows: one solves eqs.\ (\ref{kappa}) and (\ref{dkappa}) for $d$ and inserts the 
result into eq.\ (\ref{coth}). After a little algebra one arrives at eq.\ (\ref{u0eta2app}). 

The complete solution of the equations of motion is thus given by eqs.\ (\ref{ayapp}), (\ref{x4uapp}), (\ref{u0eta1app}), and (\ref{u0eta2app})
which are all given in sec.\ \ref{sec:general} of the main part of the paper.

\section{Semi-analytic solution for $T=0$ in the chirally symmetric phase}
\label{app:symmetric}

In this appendix we derive the solution (\ref{solXs}) to the equations of motion in the 
chirally symmetric phase at $T=0$.
 
Dividing eq.\ (\ref{EOMXs1}) by eq.\ (\ref{EOMXs2}) yields
\be
\frac{a_0'}{a_3'}=\frac{3b a_3 +C}{3ba_0} \, ,
\ee
where eq.\ (\ref{D0}) has been used. We write this as $3ba_0'a_0=3ba_3'a_3+Ca_3$ and integrate to obtain
\be \label{help1}
\frac{3}{2}ba_0^2 = \frac{3}{2}ba_3^2+Ca_3+K \, ,
\ee
with an integration constant $K$. This constant is easily determined by evaluating eq.\ (\ref{help1}) at $u=\infty$,
\be
K=\frac{3}{2}b\mu^2  \, .
\ee
With the new variable 
\be
z(u) = 3b\int_0^u\frac{dv}{\sqrt{v^5+b^2v^2-(3b\mu)^2+C^2}} \, ,
\ee
we can write the equations of motion (\ref{EOMXs}) as 
\begin{subequations} \label{EOMXsA}
\bea
\partial_z a_0 &=& a_3 + \frac{C}{3b} \, , \\
\partial_z a_3 &=& a_0 \, ,
\eea
\end{subequations}
which are solved by 
\begin{subequations}
\bea
a_0(z) &=& C_1\cosh z + C_2 \sinh z \, , \\
a_3(z) &=& C_1\sinh z + C_2 \cosh z -\frac{C}{3b} \, .
\eea
\end{subequations}
The integration constants $C_1$, $C_2$, $C$ can now be determined from the boundary conditions (\ref{boundaryXs}), and the resulting solutions
become
\begin{subequations} \label{solXsapp}
\bea
a_0(z) &=& \frac{\mu}{\sinh z_\infty} \sinh z \, , \\
a_3(z) &=& \frac{\mu}{\sinh z_\infty} (\cosh z - \cosh z_\infty)  \, ,
\eea
\end{subequations}
with $z_\infty\equiv z(u=\infty)$, and 
\be \label{zuapp}
z(u) = 3b\int_0^u\frac{dv}{\sqrt{v^5+b^2v^2+\frac{(3b\mu)^2}{\sinh^2 z_\infty}}} \, .
\ee
From the equation for $z(u)$ we obtain an implicit equation for $z_\infty$, 
\be \label{z8app}
z_\infty = 3b\int_0^\infty\frac{du}{\sqrt{u^5+b^2u^2+\frac{(3b\mu)^2}{\sinh^2 z_\infty}}} \, .
\ee
The complete solution is thus given by Eqs.\ (\ref{solXsapp}) with $z(u)$ given in Eq.\ (\ref{zuapp}) and $z_\infty$ to be determined
numerically from Eq.\ (\ref{z8app}).

\section{Analytic approximation for zero-temperature ``Landau level'' transition}
\label{app:bc}

Here we derive eq.\ (\ref{bcapprox}), which is an approximation to the full numerical result for the 
zero-temperature critical line within the chirally symmetric phase, see fig.\ \ref{figLLL1}. 
The free energy of the solution $z_\infty=\infty$ is
\be \label{Omega8}
\frac{\Omega_{||}(z_\infty=\infty)}{\cal N} = \int_0^\infty du\,\sqrt{u^5+b^2u^2} - \frac{3}{2}b\mu^2 \, .
\ee
This result is exact. For the nontrivial solution we use  
\be
\epsilon \equiv \frac{b}{\mu^{3/2}}
\ee
as an expansion parameter in the ansatz for $z_\infty$, 
\be \label{z8approx}
z_\infty \simeq 3Q_1^{5/2}\epsilon + Q_2\epsilon^3 \, . 
\ee
Our result will show that along the critical line $\epsilon$ is indeed small.
Inserting this ansatz into (\ref{z8}), expanding the right-hand side of this equation up to third order in $\epsilon$, and
comparing the coefficients order by order yields
\be
Q_1 = \frac{\Gamma\left(\frac{3}{10}\right)\Gamma\left(\frac{6}{5}\right)}{\sqrt{\pi}} \, , 
\qquad Q_2 = \frac{27}{4}Q_1^{15/2}-\frac{3}{2}Q_1^{9/2}\frac{\Gamma\left(\frac{9}{10}\right)\Gamma\left(\frac{3}{5}\right)}{\sqrt{\pi}} \, . 
\ee
In order to find the transition between the two solutions we need to find the zero of $\Delta\Omega$, which is the difference in the 
corresponding free energies. Inserting eq.\ (\ref{z8approx}) into the free energy (\ref{OmegaXs}), subtracting (\ref{Omega8}) from the 
result and expanding up to second order in $\epsilon$ yields
\be
\frac{\Delta \Omega}{{\cal N} \mu^{7/2}}\simeq -\frac{2}{7}\frac{1}{Q_1^{5/2}}+\frac{3}{2}\epsilon -Q_3\epsilon^2  \, ,
\ee
with 
\be \label{Q3}
Q_3 =  \frac{3}{2}Q_1^{5/2} + \frac{\Gamma\left(\frac{9}{10}\right)\Gamma\left(\frac{3}{5}\right)}{Q_1^{1/2}\sqrt{\pi}} \, .
\ee
This yields at the critical line $\Delta \Omega = 0$, 
\be
\epsilon = \frac{3}{4Q_3}-\sqrt{\frac{9}{16Q_3^2}-\frac{2}{7Q_3Q_1^{5/2}}} \simeq 0.0950977\, , 
\ee
which is the result (\ref{bcapprox}). Consequently, along the entire critical line $\epsilon$ is constant and much smaller than 1, 
which validates our approximation a posteriori.

\section{Analytic approximation for zero-temperature chiral transition at small magnetic field}
\label{app:d}
Here we derive an analytic approximation for the $T=0$ chiral phase transition line at small values of the magnetic field $b$. This is not only 
a check for our numerical result but will also help to gain further insight into IMC. 

For the chirally broken phase we first solve eqs.\ (\ref{u0eta}) up to ${\cal O}(b^2)$, 
\begin{subequations}\label{u0etaApp}
\bea
\eta &\simeq& \frac{3\ell^5\mu\,b}{16P_1^3 u_{00}} \, , \\
u_0 &\simeq& u_{00}+\frac{b^2}{8u_{00}^2} \left[\cot\frac{\pi}{16}-1-\left(\frac{3\mu}{2u_{00}}\right)^2
\left(\frac{3P_2}{P_1}-1\right)\right] \, , \label{u0etaApp2}
\eea
\end{subequations}
where the numbers $P_1$ and $P_2$ are defined in eqs.\ (\ref{P1P2}) and where $u_{00}\equiv u_0(b=0)=(2P_1)^2/\ell^2$. 
The coefficient of the $\mu^2 b^2$ term in eq.\ (\ref{u0etaApp2}) is negative which implies that $u_0$ 
may in fact decrease as a function of $b$ at fixed $\mu$. Since $u_0$ is the location of the tip of the joined D8-branes and thus is proportional 
to the constituent quark mass, IMC manifests itself not only by a symmetry restoration but also more directly by a decreasing chiral 
condensate for certain chemical potentials. This decrease is only possible in the presence of a supercurrent, i.e., an anisotropic chiral 
condensate, which introduces the $\mu$-dependence of $u_0$. (The dominant reason for IMC in the phase diagram is however the form of the free energy 
difference, as discussed in the main part of the paper, see also the following approximations.) Our numerical results show that for larger magnetic 
fields the value of $u_0$ increases again for 
all $\mu$ as expected from MC at weak coupling and converges to the value (\ref{u0B}). 

Although we are interested in the free energy difference, it is instructive to consider the free energies of broken and symmetric
phases separately. To obtain a finite result for the separate energies we subtract the vacuum contribution $\Omega_{||}(t=\mu=0)$.
Inserting eqs.\ (\ref{u0etaApp}) into the free energy (\ref{OmegaB}) yields, up to ${\cal O}(b^2)$, 
\bea
\frac{\Omega_\cup}{\mathcal{N}}&\simeq& -\frac{2}{7}\frac{P_1u_{00}^{7/2}}{2}
-b^2\,\left(\frac{P_1u_{00}^{1/2}}{2} \cot\frac{\pi}{16}+\frac{9\mu^2P_2}{8u_{00}^{3/2}}\right) \, .
\eea
For the chirally symmetric phase we distinguish between the  ``lowest Landau level'' (LLL) phase and the ``higher Landau level'' (hLL) phases
(see sec. \ref{sec:Tzero2}). The free energies of both phases are obtained from eq.\ (\ref{OmegaXs}). We find
\bea
\frac{\Omega_{||}^{\text{hLL}}}{\cal N} \simeq -\frac{2}{7}\frac{\mu^{7/2}}{Q_1^{5/2}}-Q_3\sqrt{\mu}\,b^2\, , \qquad
\frac{\Omega_{||}^{\text{LLL}}}{\cal N} =-\frac{3}{2} \mu^2b  \, ,
\eea
with $Q_1$ and $Q_3$ defined in the previous appendix. To obtain $\Omega_{||}^{\text{hLL}}$ we have used the expansion (\ref{z8approx}).

\FIGURE[t]{\label{figchiapp}
{ \centerline{\def\epsfsize#1#2{1#1}
\epsfbox{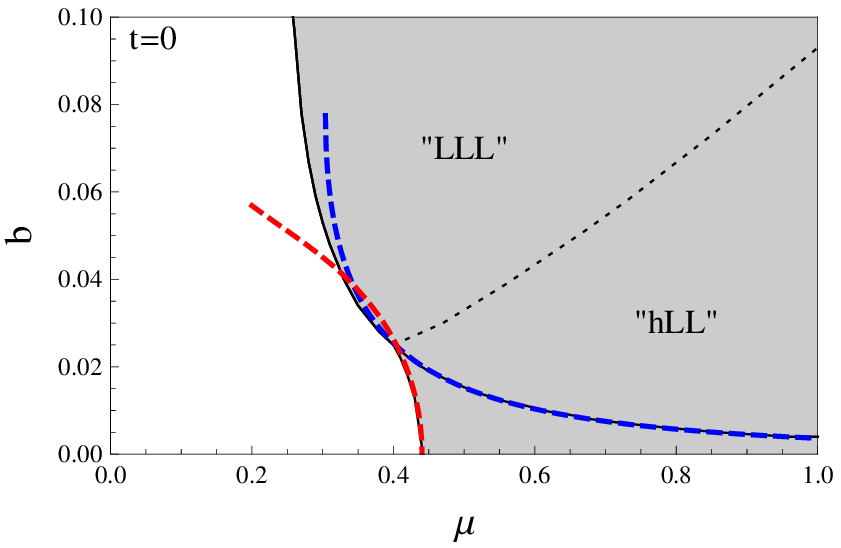}
}
}
\caption{Comparison between the analytic approximation of the chiral phase transition (dashed lines) with the 
numerical result (solid line) at zero temperature. The two dashed lines approximate the transition between the broken phase and 
the ``higher Landau level'' (hLL) phase and between the broken phase and the ``lowest Landau level'' (LLL) phase for small $b$. 
We have (unphysically) extended the phase transition line between the broken and the LLL phase into the parameter region below the (dotted) 
``Landau level'' transition line. This is firstly a check for our approximation and secondly shows that IMC is strongest in the LLL phase, with 
the ``higher Landau levels'' working against it.  
 }  
}

Consequently, the differences in free energies $\Delta\Omega = \Omega_{||}-\Omega_\cup$ 
between the broken phase and the two symmetric phases have the forms
\be
\Delta\Omega_{\rm hLL} \simeq f_1(\mu)  - f_2(\mu) b^2 \, , \qquad 
\Delta\Omega_{\rm LLL}\simeq {\rm const}-\frac{3}{2}\mu^2 b+f_3(\mu)b^2 \, , 
\ee
where the constant and the functions $f_1(\mu), f_2(\mu),f_3(\mu)$ can easily be read off from the previous equations. We plot the zeros of these free energy
differences in the $b$-$\mu$ plane and compare them to the full numerical result in fig.\ \ref{figchiapp}. 

In the relevant regime of chemical potentials we have $f_1(\mu)>0$ for $\mu<\mu_c(t=b=0)$ and $f_2(\mu)>0$ for $\mu\gtrsim 0.016/\ell^2$. 
This shows that if we start from the broken phase at $b=0$ for $0.016/\ell^2\lesssim  \mu < \mu_c(t=b=0)$ and increase $b$ at fixed $\mu$, 
chiral symmetry can be restored, provided that the sign change of $\Delta\Omega_{\rm hLL}$ occurs in a regime where our small-$b$ expansion is 
valid. The comparison with the full result shows that our expansion is indeed a very good approximation to the full result, leading to IMC in the 
hLL regime.

The transition between the chirally broken phase and the LLL phase occurs at relatively large magnetic fields where our expansion is a less 
accurate, however qualitatively still reliable, approximation. In the figure we have extended the
phase transition line between the broken and the LLL phase into the region where the hLL phase is the ground state. This extended line
is in very good agreement with our approximation and shows that IMC is strongly dominated by the LLL with the hLLs working against it. 

\providecommand{\href}[2]{#2}\begingroup\raggedright\endgroup

\end{document}